\documentclass[authoryear,preprint,review,12pt,times]{elsarticle}

\usepackage[all,dark]{draftcopy}
\usepackage{vmargin,amsmath,amssymb}
\usepackage{graphicx,subcaption}
\usepackage{epsfig}
\usepackage{float}
\usepackage{psfrag}
\usepackage[squaren]{SIunits}
\usepackage{setspace}
\usepackage{url}
\usepackage{lineno}
\usepackage{multirow}
\usepackage{color}
\usepackage{ulem}
\usepackage{rotating}

\DeclareTextCommandDefault{\trademark}{{\textcircled{R}}}

\def\Gq1{\textit{9-Grid}\,}
\def\Gq2{\textit{4.5-Grid}\,}

\newcommand{\real}{\mathbb{R}}

\newcommand{\ud}{\mathrm{d}}

\newcommand{\ppdq}[2]{ \frac{ \partial#1 }{ \partial#2 } }

\newcommand{\mynabla}{ \mathbf{ \nabla }  }
\newcommand{\myv}{ \mathbf{ u }}
\newcommand{\myu}{ \mathbf{ u }}
\newcommand{\myp}{ \mathbf{ p }}
\newcommand{\myCF}{\mathbf{F}}

\newcommand{\myn}{ \mathbf{ n } }

\newcommand{\myA}{ \text{A} }
\newcommand{\myB}{ \text{B} }
\newcommand{\myC}{ \text{C} }
\newcommand{\myM}{ \text{M} }
\newcommand{\myN}{ \text{N} }
\newcommand{\myD}{ \text{D} }
\newcommand{\myd}{ \text{d} }
\newcommand{\myG}{ \text{G} }

\newcommand{\myH}{ \text{H} }
\newcommand{\myI}{ \text{I} }

\newcommand{\myL}{ \text{L} }

\newcommand{\norm}[1]{\left\lVert#1\right\rVert}

\newcommand{\li}{ \left( }
\newcommand{\re}{ \right) }

\newcommand{\OF}{OpenFOAM}

\journal{Computer Physics Communications}
\begin{document}

\begin{frontmatter}

\title{A symmetry-preserving second-order time-accurate PISO-based method}

\author[nrg]{E.M.J. Komen\corref{cor1}}
\ead{komen@nrg.eu}
\cortext[cor1]{Corresponding author}
\author[nrg]{J.A. Hopman}
\author[nrg]{E.M.A. Frederix}
\author[terassa]{F.X. Trias}
\author[rug]{R.W.C.P. Verstappen}

\address[nrg]{Research and Innovation department, Nuclear Research and consultancy Group NRG, P.O. Box 25, 1755 ZG Petten, The Netherlands}
\address[terassa]{Heat and Mass Transfer Technological Center, Technical University of Catalonia, ESEIAAT, {c/ Colom 11, 08222} Terrassa, Spain}
\address[rug]{Bernoulli Institute for Mathematics, Computing Science, and Artificial Intelligence, University of Groningen, P.O. Box 407, 9700 AK Groningen, The Netherlands}



\begin{abstract}
A new conservative symmetry-preserving second-order time-accurate PISO-based pressure-velocity coupling for solving the incompressible Navier-Stokes equations on unstructured collocated grids is presented in this paper. This {\color{black} new} method for implicit time stepping is an extension of the conservative symmetry-preserving incremental-pressure projection method for explicit time stepping and unstructured collocated meshes of \citet{trias2014}. In order to assess and compare both methods, we have implemented them within one unified solver in the open source code {\OF}. We combine both methods with a Butcher tableau for a family of explicit and diagonally implicit Runge-Kutta temporal schemes. We assess the energy conservation properties of the implemented discretisation methods and the temporal consistency of the selected Runge-Kutta schemes using Taylor-Green vortex and lid-driven cavity flow test cases.

Although both implemented methods are based on a symmetry-preserving discretisation, we show that both methods still produce a small amount of numerical dissipation when the total pressure is directly solved from a Poisson equation. This numerical dissipation is mainly caused by the corresponding pressure error which is of $O(\Delta t \Delta h^2)$. When an incremental-pressure approach is used, where {\color{black} a} pressure correction is solved from a Poisson equation, the pressure error reduces to $O(\Delta t^2 \Delta h^2)$, yielding better conservation properties: both methods are then effectively fully-conservative. Furthermore, we conclude that all selected explicit and implicit higher order temporal schemes suffer from a reduction of the temporal order to approximately one when the pressure Poisson equation is based on the total pressure due to the presence of a pressure error of $O(\Delta t \Delta h^2)$. For the incremental pressure approach, the magnitude of the pressure error is reduced to $O(\Delta t^2 \Delta h^2)$. Consequently, second-order temporal convergence is effectively obtained when the selected higher order RK and DIRK temporal schemes are used.

\end{abstract}


\end{frontmatter}


\section{Introduction}
\label{introducton}
The conservation of mass and momentum is governed by the continuity and Navier-Stokes equations respectively. From this, it can be derived that kinetic energy is also conserved in the limit of inviscid incompressible flow. Therefore, numerical methods which conserve kinetic energy are essential for high-fidelity simulations of incompressible turbulent flows, {\color{black} see e.g.  \citet{iaccarino2010,castano2019}.} The discrete conservation properties of numerical methods for turbulent flow simulations depend strongly on how the flow variables are arranged on the grid. For the staggered grid arrangement, numerical methods which conserve both mass, momentum, and kinetic energy are published in quite a number of papers, e.g., \citet{ham2002} and \citet{morinishi2004}. However, the application of the staggered mesh approach to arbitrary unstructured grids for complex solution domains is rather complicated.

In contrast, the collocated grid arrangement has a much simpler data structure and is therefore more convenient for the treatment of complex solution domains.  This is especially true when a multigrid procedure is used \citep{pascau2011}. Furthermore, the collocated grid offers shorter computational times and reduces the required memory storage. These advantages are probably the reason why the collocated mesh approach is commonly used in commercial codes such as ANSYS-FLUENT~\citep{fluent} and STAR-CCM+~\citep{starccm} and in open source codes such as  \OF~\citep{ofoam} and Code-Saturn~\citep{archambeau2004}.

A wide stencil in the Poisson equation for the pressure occurs when central differences are used together with the collocated grid arrangement. This wide stencil causes
a decoupling of odd and even grid points which results in the well known checkerboard problem, see, e.g., \citet{versteeg2007} and \citet{ferziger1997}. \citet{rhie1983} proposed the application of the Pressure-Weighted Interpolation Method (PWIM) for the computation of the cell face velocities in order to suppress the checkerboard oscillations. Alternative velocity interpolation methods have been published in the literature since then. \citet{pascau2011} performed an extensive review of alternative velocity interpolation methods published in the literature until 2007. All considered velocity interpolation methods have in common that the face velocities are computed using a pressure gradient which is taken directly at the cell faces instead of interpolating the adjacent cell center pressure gradients. This approach yields a compact stencil in the pressure Poisson equation. This compact stencil introduces a fourth-order dissipation term in the pressure Poisson equation, which causes dissipation of kinetic energy (see, e.g., \citet{versteeg2007} and \citet{ferziger1997}).

Another problem of the collocated grid results from the fact that there are two different velocity fields in the collocated grid arrangement, namely, the convecting cell face  velocities and the cell center velocities.
The cell center velocities form the primary solution variables together with the pressure. The cell face velocities are calculated from the cell center  values of the velocity and pressure field, because the cell face velocities do not have an own transport equation. Mass conservation is applied to these cell face velocities by application of the pressure Poisson equation. \citet{morinishi1998} and \citet{feltenandlund} argue that the application of a compact stencil provides primary cell center velocities which are only approximately divergence free, which, in turn,  leads to an error in the kinetic energy, i.e., the kinetic energy is then no longer conserved.

In this paper, we focus on the error in the kinetic energy. The test cases presented in this paper clearly show that the considered energy conservation issues of the collocated grid approach correspond to numerical dissipation of kinetic energy. As shown by for example \citet{castiglioni2015}, this numerical dissipation may be larger than the Subgrid-Scale (SGS) dissipation when using explicit Large Eddy Simulation (LES), thereby making explicit LES modelling ineffective. In line with this observation, \citet{komen2017} found that explicit LES and LES without SGS closure provide rather similar results for turbulent channel flow cases when using the unstructured collocated grid approach. Given this situation, a number of papers has been published in the recent literature with the objective to reduce the numerical dissipation rate in the unstructured collocated grid solver \OF. Namely, an incompressible explicit projection method together with the classical fourth order Runge-Kutta (RK) and accelerated third order ARK3 time integration schemes have been implemented in \OF~by \citet{vuorinen2014}.
Higher order implicit RK schemes have been implemented in \OF~by \citet{Kazemi2015}, together with an incompressible iterated PISO-based procedure. In addition, they used an alternative face velocity interpolation method in order to preserve the formal order of the applied temporal schemes. 
\citet{Alessandro2018} implemented a slight variant of the approach of  \citet{Kazemi2015} in \OF. Furthermore, they implemented the incompressible explicit RK-based projection algorithm of \citet{sanderse2012}. \citet{Alessandro2018} extended both methods to flow with heat transfer. 
Together with a Butcher tableau \citep{butcher1964runge} for a family of Explicit RK (ERK) and Diagonally Implicit RK (DIRK) temporal schemes, \citet{komen2020} implemented the numerical approaches of \citet{vuorinen2014}, \citet{Kazemi2015}, and \citet{Alessandro2018} within one unified solver, named RKFoam,  based on the \OF~open source library. Using this RKFoam solver, \citet{komen2020} assessed the numerical dissipation rate of the considered numerical approaches together with a range of RK schemes. In this assessment, they also took the velocity interpolation method of \citet{rhie1983} and the standard velocity interpolation method in {\OF} into account.
\citet{komen2020} found that the pressure error, as introduced by the application of a compact stencil in the pressure Poisson equation, is the main source of numerical dissipation for all the considered methods. Furthermore, this pressure error causes a reduction of the order of accuracy of the temporal schemes to one. For compressible flows, \citet{shen2014} and \citet{modesti2017} have implemented RK-based compressible flow solvers in {\OF}.

More or less in parallel, two additional relevant papers appeared in the literature. Namely, \citet{castano2019} implemented an incremental-pressure RK4 projection method in {\OF}  and compared the performance with the standard pisoFoam solver. In the incremental-pressure method, the pressure gradient is included in the predictor step, and a pressure correction is used in the pressure Poisson equation in the projection step. From their computational results obtained for the presented inviscid Taylor-Green vortex test case, \citet{castano2019} concluded that pisoFoam clearly shows numerical dissipation, whereas their RK4 projection method is free of numerical dissipation. \citet{castano2019} prescribed the velocity boundary conditions from the analytical Taylor-Green solution, which may partially mask possible energy losses. \citet{montecchia1991} modified the velocity interpolation method of \citet{rhie1983} by introducing a scaling factor which allows to control the amount of Rhie and Chow filtering. Thereby, they reduced the amount of numerical dissipation introduced by the Rhie and Chow interpolation method. \citet{montecchia1991} performed explicit LES simulations of turbulent channel flows with this modified interpolation method. Substantial improvements were obtained by minimising the amount of numerical dissipation.

Concerning the development of fully conservative methods in collocated grid solvers, \citet{hicken2005} used so-called shift transformations which transform matrix operators for collocated variables into matrix operators for staggered variables. These shift operators were used in such a way that a fully conservative method for staggered unstructured grids was obtained for an adaptive unstructured Cartesian mesh topology. {\color{black} Given the difficulties concerning the construction of proper shift transformations for general unstructured collocated meshes \citep{vanderblij2007}, the extension of the considered approach to such meshes seems unlikely.}

\citet{iaccarino2010} presented an incompressible fully-conservative method for the Cartesian collocated grid arrangement. They use the wide stencil in the Poisson equation based on second-order central differencing.
In order to circumvent the checkerboard problem, the computed pressure solution is modified by adding a combination of null space modes in order to obtain a smooth final pressure field. We have not found a generalisation of this method to unstructured collocated grids in the literature.  As a generalisation of the symmetry-preserving discretisation method for staggered grids of \citet{verstappen2003}, \citet{trias2014} presented a fully-conservative symmetry-preserving discretisation  for incompressible flows  for the unstructured collocated mesh scheme. This discretisation method was developed for an incremental-pressure projection method for explicit time stepping. In this paper, as an extension of the method of \citet{trias2014} to implicit time stepping,
we present a new conservative symmetry-preserving second-order time-accurate PISO-based method. 
Another  second-order time-accurate PISO-algorithm for implicit time stepping has been presented by \citet{tukovic2018}. The main differences between the method of \citet{tukovic2018} and our method are: 1) a symmetry preserving discretisation, and 2) the application of a pressure correction in the pressure Poisson equation. Another important difference is the initialisation of the explicit terms in the discretised momentum equation in the predictor step at the start of each new time step. The standard approach is to initialise these terms from the cell center velocity and pressure values from the previous time step. This corresponds to a first-order extrapolation in time. In order to obtain a second-order time-accurate method, \citet{tukovic2018} use a second-order extrapolation from the cell center velocity and pressure values from the two previous time levels in the predictor step.

The main objectives of the present paper are to present our new conservative symmetry-preserving second-order time-accurate PISO-based method for implicit time stepping, and to compare the performance with the conservative symmetry-preserving projection method of \citet{trias2014} for explicit time stepping.
In order to perform this comparison, both the projection method of \citet{trias2014} and our new symmetry-preserving PISO-based method
have been implemented within one unified {\OF}-based solver. In this solver, named RKSymmFoam, we combine both implemented methods with a Butcher tableau \citep{butcher1964runge} for a family of ERK and DIRK temporal schemes.
{\color{black} The availability of implicit schemes is important for situations where explicit schemes result in too stringent stability requirements. For example, the explicit treatment of the diffusive terms in LES can result in a severe time step restriction near solid surfaces where the grid is substantially refined.
Additionally, when extending the numerical techniques to two-phase flow simulation, implicit treatment of the diffusive term becomes important for the stability of the two-phase interface.}
A secondary objective of the present paper is to provide a documentation of the applied numerical methods in the RKSymmFoam solver.
We use Taylor-Green vortex and lid-driven cavity flow test cases for the assessment and comparison of the energy conservation properties of the implemented discretisation methods and the temporal consistency of the applied Runge-Kutta schemes. The RKSymmFoam numerical methods are described in Section~\ref{sec:nmethods}.
The results obtained for the considered test cases are presented in Section~\ref{sec:RCselection}. Finally, the summary and conclusion are presented in Section~\ref{sec:conclusions}.



\section{RKSymmFoam numerical methods}
\label{sec:nmethods}

This section presents the numerical methods which constitute the basis of the RKSymmFoam solver. We use the open source library \OF5.0 \citep{ofoam}  as a platform for the development of RKSymmFoam. The main algorithm in this RKSymmFoam solver consists of the following three nested iterative levels when implicit time integration is used:
\begin{enumerate}
 \item an outer loop over each Runge-Kutta  stage $i$, indicated by $1\leq i\leq s$;
 \item an outer iteration loop for updating the non-linear convective term; 
 \item an inner PISO iteration loop for the pressure-velocity coupling.
\end{enumerate}
The goal of the PISO iteration loops \citep{Issa1982} in the third iteration level is to solve the pressure and velocity in a segregated way.
When explicit time integration is used, one projection step is used for the pressure-velocity coupling, and no outer iterations are required for updating the convective term.

The RKSymmFoam algorithm has four main control parameters which are used in order to determine the main solver settings, namely:
\begin{enumerate}
	\item the Runge-Kutta scheme, which can be either Explicit Runge-Kutta (ERK) or Diagonally Implicit Runge-Kutta (DIRK);
	\item the treatment of the non-linearity of the convective face flux when implicit time integration is used. This can either be linearised, or fully non-linear;
	\item the method used for the pressure-velocity coupling, that is, implicit PISO steps, or one explicit projection step;
    \item the treatment of the pressure, i.e., the total pressure or a pressure correction is solved for in the pressure Poisson equation.
\end{enumerate}
The RKSymmFoam algorithm and its parameters will be discussed in the subsequent sections.

\subsection{Governing equations}
\label{sec:equations}

As the starting point, the following conservation equations, which hold for a Newtonian constant-density flow without gravity and with constant physical properties, are used (\cite{Bird_etal}):
\begin{equation}\label{momentumEquationDNS}
    \ppdq{\myv}{t} =  - \mynabla \cdot ( \myv \, \myv)   -  \mynabla p +   \nu \, \mynabla^2 \, \myv,
\end{equation}
\begin{equation}\label{continuityDNS}
    \mynabla \cdot \myv  =  0,
\end{equation}
where ${\myv}$ represents the velocity, ${p}$ the pressure field divided by the constant density $\rho$, and
$\nu$ the kinematic viscosity.


\subsection{Temporal discretisation}
\label{sec:temporal}
When the spatial discretisation of section~\ref{sec:spatial} is applied to the momentum equations, the following system of ordinary differential equations is obtained
\begin{equation}
 \frac{d \myv_c}{dt} =  \myCF_c  (t, \myv_c) -  (\mynabla p)_c
 \qquad\text{with}\qquad \myv_c (t^n)=\myv_c^n,\\
\label{eq:ODEsystem}
\end{equation}
where $c$ indicates cell-centered values. Here, the pressure gradient term is not yet discretised. We use a family of ERK and  DIRK schemes for the
 discretisation of the temporal term in Eq.~\ref{eq:ODEsystem}. For both schemes, the so-called Butcher tableau \citep{butcher1964runge} is given by (see, e.g., \cite{hairer1987solving})
\begin{center} \label{tab:butcher}
    \begin{tabular}{r|lllll}
        $c_1$ & $a_{11}$ & & & & \\
        $c_2$ & $a_{21}$ & $a_{22}$ & & & \\
        $\vdots$ & $\vdots$ & $\vdots$ & $\vdots$ & & \\
        $c_s$ & $a_{s1}$ & $a_{s2}$ & $\cdots$ & $a_{ss}$ & \\
        \hline
        & $b_1$ & $b_2$ & $\cdots$ & $b_s$ \\
    \end{tabular}
\end{center}
where $a_{i\gamma}$ represent the stage weights of stage $i$, and $c_i$ are the quadrature nodes of the scheme with
\begin{equation}
    c_i = \sum_{\gamma=1}^i a_{i\gamma} \qquad\text{for}\qquad i = 1, ......, s,
\end{equation}
and $t^i = t^n+c_i \Delta t$. Furthermore, $s$ denotes the number of stages, and $b_{\gamma}$ represent the main weights of the applied Runge-Kutta scheme, with
\begin{equation}
 \sum_{\gamma=1}^s b_{\gamma} = 1.
\end{equation}
The Runge-Kutta scheme has the following form for the system of ordinary differential equations in Eq.~\ref{eq:ODEsystem},
\begin{equation}
      \myCF_c^i  =   \myCF_c(t^i, \tilde{\myv}_c^i)  \qquad\text{for}\qquad i = 1, ......, s.
      \label{RKsources}
\end{equation}
Here, the intermediate solution $\tilde{\myv}_c^i$  for stage $i$ at time $t^i$ is given by
\begin{equation}
   \tilde{ \myv}_c^{i}  =  \myv_c^n + \Delta t \li  \sum\limits^{i}_{\gamma=1} {a}_{i\gamma}  \myCF_c^{\gamma}
   - c_i(\mynabla \tilde{p} )_c^i \re,
\label{intermediateRKvelocity}
\end{equation}
and the final solution $\myv_c^{n+1}$  at time $t^{n+1} = t^n + \Delta t$ by
\begin{equation}
 \myv_c^{n+1}  = \myv_c^n + \Delta t \li  \sum\limits^{s}_{\gamma=1} b_{\gamma}  \myCF_c^{\gamma}    - (\mynabla p)_c^{n+1} \re.
\label{finalRKvelocity}
\end{equation}
The Runge-Kutta scheme consists of $s$ stages. Within each stage, a source term is determined. Each of these source terms  depends on the intermediate solution of $\tilde{\myv}_c$ at a certain stage, e.g., $\myCF_c^1$ depends on the solution for $\myCF_c$ at quadrature point $t^1$, \, $\myCF_c^2$ on the solution for $\myCF_c$ at $t^2$, etc. In the final stage, these sources are collected in order to determine the solution at $t^{n+1}$. Commonly, Runge-Kutta schemes are used as explicit schemes. The diagonal of the considered Butcher tableau is zero for all $i$ for explicit schemes.
Table~\ref{tab:RKERK} lists the explicit and implicit temporal schemes which we use in the present study.
For both explicit and implicit schemes, we select one first order scheme and two schemes of higher order.


\begin{table}[h!]
    \centering
    \caption{Explicit and implicit time integration schemes tested in this study.}
    \label{tab:RKERK}
    \begin{tabular}{llll}
        \hline
        \textbf{Scheme} & \textbf{Temporal} & \textbf{Number of} & \textbf{Ref}  \\
        & \textbf{order} & \textbf{stages}    & \\
        \hline
        {\it Explicit} & & & \\
        Forward Euler & $O(\Delta t^1)$ & 1 & \citet{ferziger1997} \\
        RK3 & $O(\Delta t^3)$ & 3 & \citet{hairer1987solving} \\
        RK4 & $O(\Delta t^4)$ & 4 &  \citet{ferziger1997} \\
        \hline
        {\it Implicit} & & & \\
        Backward Euler & $O(\Delta t^1)$ & 1 & \citet{ferziger1997} \\
        DIRK2 & $O(\Delta t^2)$ & 2 & \citet{ascher1997implicit} \\
        DIRK3 & $O(\Delta t^3)$ & 3 & \citet{ascher1997implicit} \\
    \hline
    \end{tabular}
\end{table}



\subsection{Spatial discretisation}
\label{sec:spatial}
We use a symmetry-preserving discretisation method in order to obtain a conservative finite-volume discretisation of the incompressible Navier-Stokes equations for arbitrary unstructured collocated grids.  Symmetry-preserving discretisation was first introduced by \citet{verstappen2003} for the staggered grid arrangement. \citet{trias2014} extended this discretisation method to the collocated grid arrangement. We closely follow the method of \citet{trias2014} for the spatial discretisation in this study.

\subsubsection{Discretisation of the computational domain}
\label{sec:domaindiscretization}

\begin{figure}[h]
\centering
\includegraphics[scale=0.4,angle=0]{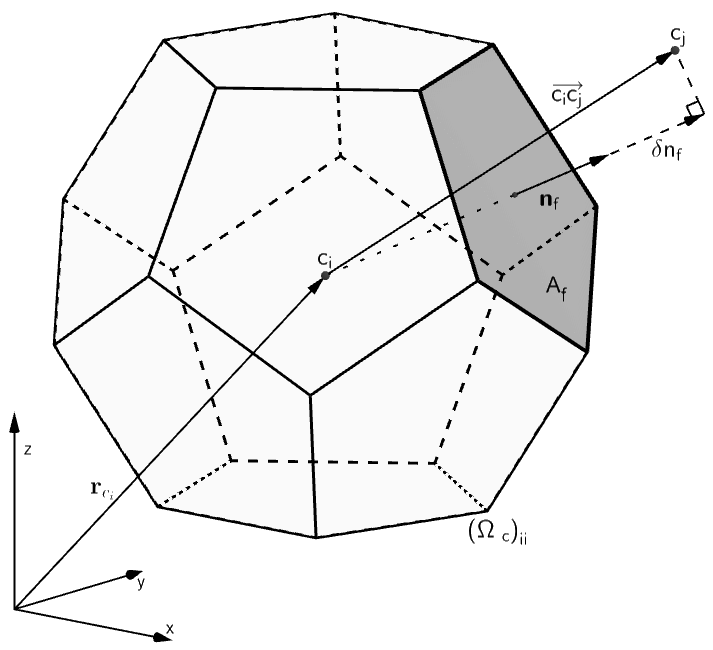}
\caption{A polyhedral control volume with volume $(\Omega_c)_{ii}$. The centroid of the control volume is
located at $c_i$. The vector $\myn_f$ is normal to the cell face to which it belongs,
and the size of $A_f$ is equal to the area of the face. $c_j$ is the cell
center of the neighbouring cell.}
\label{fig:polyhedral_cell}
\end{figure}

Within \OF, the equations are solved in a fixed Cartesian coordinate system.  The computational domain is subdivided in a finite number of control volumes.
These control volumes, or cells, can be of a general polyhedral shape with a variable number of neighbouring cells, thereby creating an arbitrary unstructured mesh. All faces of the control volumes are flat, and point $c_i$ is located at the centroid of cell $i$. 
Cell $i$ has a volume of $(\Omega_c)_{ii}$. Figure~\ref{fig:polyhedral_cell} shows polyhedral cell $i$ with its centroid $c_i$, and face $f$ with area $A_f$ and unit normal vector $\myn_f$. Furthermore, Fig.~~\ref{fig:polyhedral_cell} shows the centroid $c_j$ of a neighbouring cell $j$ which shares face $f$ with $i$. The vector connecting the centroids of cell $i$ and $j$ is represented by $\overrightarrow{c_i c_j}$. The projection of the distance between centroids $i$ and $j$ normal to face $f$ is prepresented by $\delta n_f = |\myn_f \cdot \overrightarrow{c_i c_j}|$.

The mesh consists in total of $n$ computational cells and $m$ corresponding cell faces.
The block diagonal matrix $\Omega_c \in \real^{3n \times 3n}$ is given by $ \Omega_c = \myI_3 \otimes \Omega_{c\#} $,
where $\Omega_{c\#} \in \real^{n \times n}$ is a diagonal matrix containing the cell-centered control volumes $(\Omega_c)_{ii}$, $\myI_3 \in \real^{3 \times 3}$ the identity matrix, and $\otimes$ the Kronecker product.
The diagonal matrix $\Omega_{s} \in \real^{m \times m}$ consists of the staggered control volumes $(\Omega_s)_{ff} = \delta n_f A_f$.

\subsubsection{Definition of the collocated operators}
\label{sec:operator_definition}
Application of a finite-volume spatial discretisation to Eqs.~\ref{momentumEquationDNS} and \ref{continuityDNS} yields the following equations in a matrix-vector notation for an arbitrary unstructured collocated mesh
\begin{equation}
    \label{eq:NSDiscMom}
    \Omega_c \frac{\ud\myu_c}{\ud t} + \myC(\myu_s)\myu_c + \myD \myu_c + \Omega_c \myG_c \myp_c = \mathbf{0}_{c,3n},
\end{equation}
\begin{equation}
    \label{eq:NSDiscComp}
    \myM \myu_s=\mathbf{0}_{c,n},
\end{equation}
where $\myp_c = ( p_1 , p_2, \dots \vphantom{f^f_f} , p_n )^T \in \real^{n}$ is the cell-centered pressure and $\myu_c \in \real^{3n}$ the cell-centered velocity. The subscripts $c$ and $s$ indicate whether the variables are cell-centered or staggered at the faces.
The cell-centered velocity $\myu_c$ is arranged as $\myu_c = ( \myu_1 , \myu_2 , \myu_3 )^T$, where $\myu_i = \left( {(u_i)}_{1} , {(u_i)}_{2}, \dots \vphantom{f^f_f}, {(u_i)}_{n} \right)^T$ are the vectors containing the velocity components corresponding to the $x_i$-direction. The auxiliary staggered face normal velocity field is represented by $\myu_s = \left( {(u_s)}_{1}, {(u_s)}_{2}, {(u_s)}_{3}, \dots \vphantom{f^f_f}, {(u_s)}_{m} \right)^T \in \real^{m}$. The null-vector $\mathbf{0}_{c,3n} \in \real^{3n}$, whereas $\mathbf{0}_{c,n} \in \real^{n}$.
The cell-centered and staggered velocities are related via the linear shift transformation  $\Gamma_{c \rightarrow s} \in \real^{m \times 3n}$, that is,
\begin{equation}
\label{eq:usdef}
    \myu_s \equiv \Gamma_{c \rightarrow s} \myu_c.
\end{equation}
This shift transformation consists of an interpolation and a projection to the face normal operator, that is,
\begin{equation}
\label{eq:shifttransformation}
     \Gamma_{c \rightarrow s} = \myN_s \Pi,
\end{equation}
where the matrices $\myN_s \in \real^{m \times 3m}$ and $\Pi \in \real^{3m \times 3n}$ are given by respectively
\begin{equation}
\myN_s = \left(
\begin{array}{ccc}
\myN_{s,1} & \myN_{s,2} & \myN_{s,3}
\end{array}
\right) \hspace{5mm} \text{and} \hspace{5mm}
\Pi = \myI_3 \otimes \Pi_{c \rightarrow s},
\end{equation}
where $\myN_{s,i} \in \real^{m \times m}$ are the diagonal matrices which contain the $x_{i}$-components of the
face normal vectors, and $\Pi_{c \rightarrow s} \in \real^{m \times n}$ is the operator that interpolates cell-centered values to face values. The discussion on the discretisation of this operator follows later.

The block diagonal matrices $\myC(\myu_s)$ and $\myD$ belong to $\real^{3n \times 3n}$ and are given by respectively
\begin{equation}
\label{convdiff}
 \myC(\myu_s) = \myI_3 \otimes \myC_{c\#}(\myu_s), \hspace{5mm} \myD = \myI_3 \otimes \myD_{c\#},
\end{equation}
$\myC_{c\#}(\myu_s) \in \real^{n \times n}$ and $\myD_{c\#} \in \real^{n \times n}$ are the cell-centered convective and diffusive operators of a discrete scalar field. $\myG_c \in \real^{3 n \times n}$ denotes the discrete gradient operator, and $\myM \in \real^{n \times m}$ the face-to-center discrete divergence operator.

\subsubsection{Symmetry requirements on the collocated operators}
\label{sec:operator_requirements}
There is a direct relationship between the conservative properties of Eqs.~\ref{momentumEquationDNS} and \ref{continuityDNS} and the symmetries of their differential operators \citep{verstappen2003}. In order the retain these conservative properties, it is required that the discrete operators preserve the symmetries of their continuous analogs. In this section, the symmetry requirements for the discrete operators will be determined, with the goal to obtain spatial discretisations which exactly conserve the total kinetic energy for inviscid flows. The conservation equation of the total discrete kinetic energy $\norm{\myu_c}^2 = \myu_{c}^T\Omega_c \myu_c$ can be obtained by left-multiplying equation \ref{eq:NSDiscMom} by $\myu_c^T$ and summing the resulting expression with its transpose, i.e.,
\begin{equation}
    \label{eq:KinEnEvolution}
    \frac{\ud}{\ud t}\norm{\myu_c}^2 = -\myu_c^T\big(\myC(\myu_s)+\myC^T(\myu_s)\big)\myu_c - \myu_c^T\big(\myD+\myD^T\big)\myu_c - \myu_c^T\Omega_c \myG_c \myp_c - \myp_c^T \myG_c^T\Omega_c^T\myu_c.
\end{equation}
For inviscid flows, where $\nu = 0$, the total discrete kinetic energy $\norm{\myu_c}^2$ is conserved if both the convective and pressure terms vanish in Eq.~\ref{eq:KinEnEvolution}, that is,
\begin{equation}
    \label{eq:EconserveC}
    \myu_c^T\big(\myC(\myu_s)+\myC^T(\myu_s)\big)\myu_c = 0,
\end{equation}
\begin{equation}
    \label{eq:EconserveP}
    \myu_c^T\Omega_c \myG_c \myp_c + \myp_c^T \myG_c^T\Omega_c^T \myu_c = 0.
\end{equation}
Equation \ref{eq:EconserveC} implies that
\begin{equation}
\label{eq:CSkewSym}
    \myC(\myu_s) = -\myC^T(\myu_s),
\end{equation}
which means that the operator $\myC$ should be skew-symmetric in order to conserve the total discrete kinetic energy. Equation~\ref{eq:EconserveP} is fulfilled when the following condition is met
\begin{equation}
    \label{eq:OmegacGcMGammacs}
    -(\Omega_c \myG_c)^T = \myM\Gamma_{c \rightarrow s}.
\end{equation}
Therefore, when the conditions in Eqs.~\ref{eq:CSkewSym} and \ref{eq:OmegacGcMGammacs} are met, the total discrete kinetic energy equation reduces to
\begin{equation}
    \frac{d}{dt}\norm{\myu_c}^2 = -\myu_c^T(\myD+\myD^T)\myu_c \leq 0.
\end{equation}
The condition that the time rate of change of the total discrete kinetic energy is less than or equal to zero follows from the condition that the diffusion term should be strictly dissipative. Therefore, $(\myD+\myD^T)$ must be positive-definite. Moreover, although not strictly necessary, $\myD$ is assumed to be symmetric, like its continuous analog $-\Delta$.

\subsubsection{Pressure-velocity coupling for explicit time integration}

\label{sec:pvcoupling}
The explicit pressure-velocity coupling method which we use in this study is explained for the forward Euler time integration scheme, since higher order explicit Runge-Kutta integration schemes essentially consist of a sequence of forward Euler stages. An intermediate velocity field $\myv_c^p$ is computed from the following predictor step
\begin{equation}
\label{pc1_step1}
    \frac{\myv_c^{p}-\myv_c^n}{\Delta t^n} = \myH(\myv_s^n) \myv_c^n - \myG_c \myp_c^{p},
\end{equation}
where $\myH(\myv_s^n) \myv_c^n \equiv -\Omega_c^{-1}(\myC(\myv_s^n) \myv_c^n + \myD \myv_c^n)$, with $\myH(\myv_s^n) \in \real^{3n \times 3n}$, and $\Delta t^n$ is the time between time levels $n$ and step $n+1$. In the classical first-order time accurate Chorin projection method \citep{chorin1968}, $\myp_c^{p} = \mathbf{0}_{c,n}$, whereas $\myp_c^{p} = \myp_c^{n}$ corresponds to the second-order time accurate projection method of \citet{Kan1986}.
Since the intermediate velocity field $\myv_c^p$ is not divergence free, the final values for time step $n+1$ are obtained by adding the following corrections to the intermediate values: $\myv_c^{n+1}  =  \myv_c^p + \myv_c' $ and $\myp_c^{n+1}  =  \myp_c^p + \myp_c' $.
The following relation holds between the velocity and pressure corrections $\myv_c'$ and $\myp_c'$
\begin{equation}\label{pc1_step1_corr}
\myv_c' =   - \Delta t \myG_c \myp_c'.
\end{equation}
Taking the divergence, this yields
\begin{equation}\label{pc1_step1_corr2}
\myM \Gamma_{c \rightarrow s} \myv_c' =   - \Delta t \myM \Gamma_{c \rightarrow s} \myG_c \myp_c'.
\end{equation}
The gradient operator $\myG_c$ and the center-to-face staggered gradient operator $\myG_s \in \real^{m \times n}$ are related via the linear shift transformation  $\Gamma_{s \rightarrow c} \in \real^{3n \times m}$, that is,
\begin{equation}
\label{eq:Gcdef}
    \myG_c \equiv \Gamma_{s \rightarrow c} \myG_s,
\end{equation}
where the staggered gradient operator $\myG_s$ is related as follows to the discrete divergence operator $\myM$ 
\begin{equation}
\label{eq:GradsDef}
    \myG_s \equiv -\Omega_s^{-1} \myM^T.
\end{equation}

In order to guarantee that the contribution of the pressure gradient term to the global kinetic energy vanishes, a constraint has to be imposed on the shift operators $\Gamma_{c\rightarrow s}$ and $\Gamma_{s\rightarrow c}$. This constraint is obtained as follows: first, the following expression for the cell-centered discrete gradient operator $\myG_c$ is obtained from the definition of the staggered discrete gradient operator $\myG_s$ in Eq.~\ref{eq:GradsDef}
\begin{equation}
\label{eq:GradcDef2}
    \myG_c = - \Gamma_{s\rightarrow c} \Omega_s^{-1} \myM^T.
\end{equation}
Subsequently, the constraint for the shift operators is obtained using the requirement in Eq.~\ref{eq:OmegacGcMGammacs},
\begin{equation}
\label{eq:shiftoperatorconstraint}
    \li \Omega_c \Gamma_{s\rightarrow c} \Omega_s^{-1} \myM^T \re^T = \myM \Gamma_{c\rightarrow s}
    \,\,\,\,\, \rightarrow \,\,\,\,\,
    \Gamma_{s\rightarrow c} = \Omega_c^{-1} \Gamma^T_{c\rightarrow s} \Omega_s.
\end{equation}
Using the expressions in Eqs.~\ref{eq:GradcDef2} and \ref{eq:shiftoperatorconstraint},
and $\myM \myu_s' =  \myM (\myu_s^{n+1} - \myu_s^p) = - \myM  \myu_s^p$,
the following pressure Poisson equation is obtained from Eq.~\ref{pc1_step1_corr2}
\begin{equation}\label{pc1_step1_Poisson2}
\myM \Gamma_{c\rightarrow s} \myG_c \myp_c' =
- \myM \Gamma_{c\rightarrow s} \Omega_c^{-1} \Gamma^T_{c\rightarrow s} \myM^T \myp_c'
 = \frac{1}{\Delta t} \myM \myu_s^p,
\end{equation}
where the Laplacian $\myL^w \equiv - \myM \Gamma_{c\rightarrow s} \Omega_c^{-1} \Gamma^T_{c\rightarrow s} \myM^T$ can be introduced in the pressure Poisson Eq.~\ref{pc1_step1_Poisson2}, where $\myL^w \in \real^{n \times n}$ is a symmetric negative-definite matrix. This Laplacian $L^w$ assures that the pressure gradient contribution to the global kinetic energy vanishes. However, this Laplacian is based on a wide stencil, which is not possible for the pressure Poisson equation in the present \OF~data structure. Furthermore, the application of this wide stencil introduces the checkerboard problem. In order to arrive at a Laplacian based on a compact stencil, the approximation $\Gamma_{c \rightarrow s} \myG_c = \Gamma_{c \rightarrow s} \Gamma_{s \rightarrow c} \myG_s \approx \myG_s$ is made. This way, the two interpolation steps present in the wide Laplacian $\myL^w$ are eliminated.
As a result, the following Poisson equation for the pressure correction $\myp_c'$ is obtained from Eq.~\ref{pc1_step1_corr2}
\begin{equation}\label{pc1_step1_Poisson}
\myM \myG_s \myp_c' =  - \myM \Omega_s^{-1} \myM^T \myp_c' = \frac{1}{\Delta t} \myM \myu_s^p.
\end{equation}
Now, the compact Laplacian $\myL \equiv  - \myM \Omega_s^{-1} \myM^T$ can be introduced in the pressure correction Eq.~\ref{pc1_step1_Poisson}, where $\myL \in \real^{n \times n}$ is a also symmetric negative-definite matrix. This Laplacian operator inherits the boundary conditions from the divergence operator $\myM$. Therefore, no additional boundary conditions are needed \citep{trias2014}.
It is noted that $\Gamma_{c \rightarrow s} \Gamma_{s \rightarrow c} = \myI$ holds in general only approximately.
Consequently, the velocity is approximately divergence-free and
the application of the compact Laplacian $L$ does in general not guarantee that the pressure contribution to the global kinetic energy vanishes.

Once the pressure correction $\myp'$  is obtained from the Poisson equation Eq.~\ref{pc1_step1_Poisson}, the velocity correction $\myv'$ can be calculated using Eq.~\ref{pc1_step1_corr}. Finally, the new velocity and pressure fields at the next time step can be calculated using the velocity and pressure corrections $\myv_c'$ and $\myp_c'$.

In the \citet{Kan1986} projection method, the pressure correction $\myp_c'$ in $\myp_c^{n+1}  =  \myp_c^p + \myp_c' $ {\color{black} with $\myp_c^p = \myp_c^n$} is of $O(\Delta t)$. In this study, we also analysed the application of $\myp_c^{p} = 2 \myp_c^{n} - \myp_c^{n-1}$, where the pressure correction $\myp_c'$ is of $O(\Delta t^2)$.


\subsubsection{Pressure-velocity coupling for implicit time integration}
For implicit time integration, we use the PISO (Pressure-Implicit with Splitting of Operators) approach of \citet{Issa1982} as a base for the pressure-velocity coupling in our RKSymmFoam solver. Similar as for explicit time integration, we explain the PISO approach for the implicit backward Euler time integration scheme, since higher order implicit Runge-Kutta integration schemes essentially consist of a sequence of backward Euler stages. The PISO method consists of one predictor step followed by $m_{corr}$ corrector steps (or: inner iterations). 
PISO proceeds according to the following steps \citep{Issa1982}:

1) {\it Predictor step}:
The first intermediate velocity field $\myv_c^{1}$ is computed from the following predictor equation
\begin{equation}
\label{pc5_step1_mom}
    \frac{\myv_c^{1}-\myv_c^n}{\Delta t^n} = \myH(\myv_s^p) \myv_c^p - \myG_c \myp_c^{p},
\end{equation}
where $\myH(\myv_s^p) \myv_c^p \equiv -\Omega_c^{-1}(\myC(\myv_s^p) \myv_c^p + \myD \myv_c^p)$.
In the original PISO method \citep{Issa1982}, the pressure at the old time level $n$ is taken for $\myp_c^{p}$ i.e., $\myp_c^{p} = \myp_c^{n}$, and $\myH(\myv_s^p) \myv_c^p = \myH(\myv_s^1) \myv_c^1 \approx \myH(\myv_s^n) \myv_c^1$, where $\myH(\myv_s^n)$ remains constant during the inner PISO iterations. In our RKSymmFoam solver, we use $\myH(\myv_s^p) \myv_c^p = \myH(\myv_s^n) \myv_c^n$ in order to avoid implicit treatment of the convective and diffusive terms. 
The obtained first intermediate velocity field $\myv_c^{1}$ will generally not be divergence free. Therefore, we perform subsequent corrector steps.

2) {\it corrector steps}:
In each corrector step, a new pressure field $\myp_c^{k}$ and a corresponding revised velocity field $\myv_c^{k+1}$ which is divergence free, that is, $\myM\Gamma_{c \rightarrow s} \myu_c^{k+1} = \mathbf{0}_{c,n},$ are determined. The momentum equation in the corrector steps is considered in the following form, where the velocities are treated explicitly
\begin{equation}
\label{pc5_step2_mom}
    \frac{\myv_c^{k+1}-\myv_c^n}{\Delta t^n} = \myH(\myv_s^n) \myv_c^k - \myG_c \myp_c^{k}, \qquad \mbox{with} \qquad k = 1, 2,\dots,m_{corr}.
\end{equation}
To improve the stability of this equation, the operator $\myH(\myv_s^n)$ is split in a diagonal part $\myA^d (\myv_s^n)$ which operates on $\myv_c^{k+1}$ and an off-diagonal part $\myH^{od}(\myv_s^n)$ which operates on $\myv_c^{k}$.
As a result, Eq.~\ref{pc5_step2_mom} can be recast in the following form
\begin{equation}
    \myB \myu_c^{k+1} = \Delta t \myH^{od}(\myv_s^n) \myv_c^k + \myv_c^n - \Delta t \myG_c \myp_c^{k}, \qquad \mbox{with} \qquad k = 1, 2,\dots,m_{corr},
\end{equation}
where $\myB = \myI-\Delta t^n \myA^d (\myv_s^n) \in \real^{3n \times 3n}$. Therefore, the following equation can be obtained for $\myu_c^{k+1}$
\begin{equation}
\label{eq:ImpUc}
    \myu_c^{k+1} = \myB^{-1}(\Delta t \myH^{od}(\myv_s^n) \myv_c^k + \myu_c^n) - \Delta t \myB^{-1}\myG_c \myp_c^{k}, \qquad \mbox{with} \qquad k = 1, 2,\dots,m_{corr}.
\end{equation}
A Poisson equation for the pressure $\myp_c^{k}$ can be obtained by taking the divergence of Eq.~\ref{eq:ImpUc} and by using $\myM\Gamma_{c\rightarrow s}\myu_c^{k+1} = \mathbf{0}_{c,n}$, that is
$$
   \myM \Gamma_{c\rightarrow s} \myB^{-1} \myG_c \myp_c^{k} =
    \myM \Gamma_{c \rightarrow s} \myB^{-1} \Gamma_{s\rightarrow c} \myG_s \myp_c^{k} =
    \myM \myB_s^{-1} \Omega_s^{-1} \myM^T \myp_c^{k} =
$$
\begin{equation}
\label{eq:ImpPoisson}
   \frac{1}{\Delta t} \myM\Gamma_{c\rightarrow s}\myB^{-1}(\Delta t\myH^{od}(\myv_s^n) \myu_c^k + \myu_c^n),
    \qquad \mbox{with} \qquad k = 1, 2,\dots,m_{corr},
\end{equation}
with $ \Gamma_{c\rightarrow s} \myB^{-1} \Gamma_{s\rightarrow c} \approx \myB_s^{-1}$,
where $\myB_s \in \real^{m \times m}$ is a square diagonal matrix. That is, the sequence of interpolation of face values to cell center values, multiplication of the resulting cell center values with the diagonal components of $\myB^{-1}$, followed by an interpolation of the obtained cell center values to the face again is replaced by a multiplication of the face values with the diagonal components of $\myB_s^{-1}$. As a result, in contrast with the  Laplacian $\myL =  - \myM \Omega_s^{-1} \myM^T$ for the explicit case, the Laplacian $\myM \myB_s^{-1} \Omega_s^{-1} \myM^T$ for the implicit case is multiplied with the diagonal components of $\myB_s^{-1}$.
Similar as for the explicit case, the Laplacian for the implicit case has a compact stencil.
The diagonal components $\mathrm{diag}(\myB_s) \in \real^m$ of $\myB_s$ can be obtained by interpolating the cell center values of the diagonal of $\myB_\#$ to the face, that is,
\begin{equation}
\label{eq:diogonalB}
   \mathrm{diag}(\myB_s) = \Pi_{c\rightarrow s} \mathrm{diag}(\myB_\#),
\end{equation}
with $\myB = \myI_3 \otimes \myB_{\#}$, $\mathrm{diag}(\myB_\#) \in \real^n$, and $\Pi_{c\rightarrow s} \in \real^{m \times n}$.

Once the new pressure field $\myp_c^{k}$ is obtained from the Poisson equation Eq.~\ref{eq:ImpPoisson}, the corresponding revised velocity field $\myv_c^{k+1}$,  which satisfies $\myM\Gamma_{c \rightarrow s} \myu_c^{k+1} = \mathbf{0}_{c,n}$, can be computed from Eq.~\ref{eq:ImpUc}. This process is repeated for $m_{corr}$ iterations, until an inner iteration criterion is satisfied. Then, the inner PISO iteration process is finalised by updating the velocity which is used to resolve the non-linearity in the convective flux term in the operator $\myH$ with the new velocity  $\myv_s^{m_{corr}}$ for the next outer iteration.

This completes the description of the iterative PISO approach for application to the symmetry-preserving discretisation method used in this paper. Similar to the standard PISO formulation of \citet{Issa1982}, the Poisson equation is based on $\myp_c^{k}$. In this work, we also use a modified PISO approach based on the pressure correction ${\myp'}_c^k$, with $\myp_c^{k} =  \myp_c^{p} + {\myp'}_c^k$. This pressure correction ${\myp'}_c^k$ is obtained from the following Poisson equation
\begin{equation}
\label{eq:ImpPoisson2}
    \myM \myB_s^{-1} \Omega_s^{-1} \myM^T {\myp'}_c^k =
   \frac{1}{\Delta t} \myM\Gamma_{c\rightarrow s}\myB^{-1}(\Delta t\myH^{od}(\myv_s^n) \myu_c^k + \myu_c^n - \myG_c \myp_c^{p}),
     \qquad \mbox{with} \qquad k = 1, 2,\dots,m_{corr}.
\end{equation}
For $\myp_c^{p}$ in Eq.~\ref{eq:ImpPoisson2}, we use respectively $\myp_c^{p} = \myp_c^{n}$ and $\myp_c^{p} = 2 \myp_c^{n} - \myp_c^{n-1}$ in this study.
Note that the cell-centered gradient of $\myp_c^{p}$ is taken explicitly. Subsequently, after multiplication with the diagonal components of $\myB^{-1}$, this cell-centered gradient is interpolated to the faces. Taking the divergence of the obtained interpolated pressure gradients provides a wide stencil for the Laplacian of $\myp_c^{p}$. Consequently, a larger part of the total pressure is evaluated on a wide stencil, and only the Laplacian of the pressure correction is based on a compact stencil. A a result, the pressure gradient contribution to the kinetic energy becomes smaller, and kinetic energy is better conserved. {\color{black} Namely, as will be demonstrated in section~3, the kinetic energy dissipation error caused by the application of the compact Laplacian is of $O(\Delta t \Delta h^2)$  when the compact Laplacian is based on the total pressure $\myp_c$ ($\myp_c^{p} = \bf{0}_c$) and of $O(\Delta t^2 \Delta h^2)$  when it is based on the pressure correction $\myp'_c$ with $\myp_c^{p} = \myp_c^{n}$.}

\subsubsection{Discretisation of the operators}
The discretisation of the operators defined in the preceding sections is selected such that the (skew-)symmetry properties of the operators are preserved. These requirements restrict the local truncation error in some cases.

{\it Divergence, gradient, and Laplacian operators}\\
Integration of the conservation of mass equation (Eq.~\ref{continuityDNS}) over cell $i$ with volume $(\Omega_c)_{ii}$ yields
\begin{equation}\label{continuityOperator1}
  \int_{(\Omega_c)_{ii}}  \mynabla \cdot \myv \, \myd V =  \,\,
   \int_{\partial (\Omega_c)_{ii}}  \myv \cdot \myn \, \myd S = \,\,
   \sum_{f\in F_f(i)} \int_{S_f}  \myv \cdot \myn \, \myd S,
\end{equation}
where $F_f(i)$ is the set of faces bordering cell $i$. When the discrete normal velocities $\left[ \myu_s \right]_f$ are located at the centroid of the cell faces $f$, a second-order accurate discretisation of Eq.~\ref{continuityOperator1} is obtained as follows
\begin{equation}\label{continuityOperator2}
   \sum_{f\in F_f(i)} \int_{S_f}  \myv \cdot \myn \, \myd S \approx
   \sum_{f\in F_f(i)} \left[ \myu_s \right]_f A_f.
\end{equation}
{\color{black} Note that $\left[ \myu_s \right]_f$ denotes the scalar value of $\myu_s$ at face $f$.}
Now, the divergence operator $\myM \in \real^{n \times m}$ is defined as
\begin{equation}\label{discretedivergence}
   \left[ \myM \myu_s  \right]_i \equiv
   \sum_{f\in F_f(i)} \left[ \myu_s \right]_f A_f = 0.
\end{equation}
The divergence operator acts on the outward face normal velocities. The matrix entries $[\myM]_{if}$ have non-zero values when face $f$ lies between cell $i$ and neighbour cell $j$. These non-zero entries are given by: $[\myM]_{if} = A_f$ when $i < j$ and $[\myM]_{if} = - A_f$ when $i > j$.

Using this definition of $[\myM]_{if}$ and $(\Omega_s)_{ff} = \delta n_f A_f$, and Eq.~\ref{eq:GradsDef}, the center-to-face staggered gradient operator $\myG_s \in \real^{m \times n}$ is constructed.
The matrix has non-zero entries $[\myG_s]_{fi}$ when face $f$ lies between $i$ and neighbour cell $j$. Its entries are given by: $[\myG_s]_{fi} = - 1 /{\delta n_f}$ when $i < j$ and $[\myG_s]_{fi} =  1 /{\delta n_f}$ when $i > j$.
The discretisation of the pressure gradient at face $f$ is therefore obtained from
\begin{equation}\label{discretedivergence2}
[\myG_s \myp_c]_{f} =  \frac{p_{c_j}- p_{c_i}}{\delta n_f}.
\end{equation}
Since $\Omega_s \myG_s$ is equal to $-\myM^T$ (from Eq.~\ref{eq:GradsDef}), the expression in Eq.~\ref{discretedivergence2} can also be obtained directly from Eq.~\ref{discretedivergence}.
Finally, the Laplacian operator is computed according to $\myL =  - \myM \Omega_s^{-1} \myM^T$ for explicit time integration and $\myL = \myM \myB_s^{-1} \Omega_s^{-1} \myM^T$ for implicit time integration.

{\it Convective and diffusive operators}\\
{\color{black} The convective term for cell $i$, denoted by scalar value $\left[ \myC(\myu_s)\myu_c \right]_i$,} is equal to the sum of the product of the flux at each face of cell $i$ and the corresponding interpolated velocity at each face. That is, the matrix entries $[\myC(\myu_s)]_{ij}$  contain the flux from cell $i$ to cell $j$ at face $f$,  multiplied by the interpolation weight $w_{jf}$ which is used to interpolate $[\myu_c]_j$ to face $f$. {\color{black} Note that, in contrast to $[\myu_s]_f, \, [\myu_c]_j$ denotes a vector which contains the x-, y-, and z-components of $u_c$ at cell $j$.}
Analogously, matrix entry $[\myC(\myu_s)]_{ji}$ contains the flux from cell $j$ to cell $i$ at face $f$ multiplied by interpolation weight $w_{if}$ with which $[\myu_c]_i$ is interpolated to face $f$.
That is, the off-diagonal terms of $[\myC(\myu_s)]$ are non-zero when $i$ and $j$ are neighbouring cells.
In order to preserve global kinetic energy, $\myC(\myu_s)$ has to be skew-symmetric, i.e., $[\myC(\myu_s)]_{ij} = - [\myC(\myu_s)]_{ji}$. Therefore, the interpolation weight $w_{if}$ equals $w_{jf}$. Since $w_{if} + w_{jf} = 1$, skew-symmetry of $\myC(\myu_s)$ is only preserved when interpolation from cell center to face values occurs with interpolation weights of $\frac{1}{2}$, which corresponds to midpoint interpolation. Therefore, the linear shift transformation  $\Gamma_{c \rightarrow s} = \myN_s \Pi  $, which  consists of an interpolation and a projection to the face normal operator, is discretised as follows
\begin{equation}
\label{eq:discreteshifttransformation}
    \left[ \myu_s \right]_f =  \left[ \Gamma_{c \rightarrow s} \myu_c \right]_f = \frac{1}{2} \li \left[\myu_c\right]_{i} + \left[\myu_c\right]_{j} \re \cdot \myn_f,
\end{equation}
and the matrix entries $ \left[ \myC(\myu_s) \right]_{ij}$ with $i \neq j$ are equal to half of the flux through the faces $f$ between neighbouring cells $i$ and $j$, i.e.,
\begin{equation}
\label{eq:convectiveentryij}
  \left[ \myC(\myu_s) \right]_{ij}  = \frac{1}{2} \left[ \myu_s \right]_f A_f.
\end{equation}
Furthermore, for skew-symmetry of $\myC(\myu_s)$, the diagonal elements must be zero, that is,
\begin{equation}
\label{eq:zero diagonal convective}
  \left[ \myC(\myu_s) \right]_{ii}  = \frac{1}{2} \sum_{f\in F_f(i)}  \left[ \myu_s \right]_f A_f = 0.
\end{equation}
This condition is fulfilled because of Eq.~\ref{discretedivergence}, that is, the discrete divergence of $\myu_s$ vanishes.
Midpoint interpolation assures that the off-diagonal elements of the matrix $\myC(\myu_s)$ satisfy the skew-symmetric condition too.

Finally, when the discrete normal velocities $\left[ \myu_s \right]_f$ are located at the centroid of the cell faces $f$, a second-order accurate discretisation of the convective term for cell $i$ is obtained as follows
\begin{equation}
\label{eq:discreteconvective term}
  \left[ \myC(\myu_s)\myu_c \right]_i = \sum_{f\in F_f(i)} \frac{1}{2} \li \left[\myu_c\right]_{i} + \left[\myu_c\right]_{j} \re
    \left[ \myu_s \right]_f A_f,
\end{equation}

The diffusion term  $\left[ \myD \myu_c \right]_i$ for cell $i$ is equal to the sum of the diffusion fluxes of $\myu_c$ at each face of cell $i$. The off-diagonal terms are non-zero for matrix entries $[\myD]_{ij}$ when cell $i$ and cell $j$ are neighbours. These off-diagonal terms are given by
\begin{equation}
\label{eq:diffusiveentryij}
  \left[ \myD) \right]_{ij}  = - \frac{\nu A_f}{\delta n_f},
\end{equation}
where $\nu$ is the kinematic viscosity. The diagonal terms are given by the negative of the sum of the off-diagonal terms, i.e.,
\begin{equation}
    \left[ \myD) \right]_{ii} = -\sum_{j=1}^{i-1}\left[ \myD) \right]_{i,j}-\sum_{j=i+1}^{n}\left[ \myD) \right]_{i,j}
\end{equation}
Notice that $\left[ \myD) \right]_{i,j} = \left[ \myD) \right]_{j,i}$ and $\left[ \myD) \right]_{i,j} < 0$. Therefore $\left[ \myD) \right]_{ii} > 0$, and $\left[ \myD) \right]$ is symmetric and positive definite.
The overall diffusion term  $\left[ \myD \myu_c \right]_i$ for cell $i$ is obtained from
\begin{equation}
\label{eq:diffusiontermcelli}
  \left[ \myD \myu_c \right]_i  =  \sum_{f\in F_f(i)}   \frac{\nu A_f}{\delta n_f} \li [\myu_c]_j - [\myu_c]_i \re.
\end{equation}
With $[\myM]_{if} = A_f, \, (\Omega_s)_{ff} = \delta n_f A_f$, and $\myD_{c\#} = - \nu \myL =  \nu \myM \Omega_s^{-1} \myM^T$, the considered terms of the diffusion matrix can also be obtained.


\section{Results for basic test cases}
\label{sec:RCselection}
In this section, we first explain the different possible error sources which are present in the applied RKSymmFoam solver. Subsequently, we present an analysis of the effect of the application of the different choices for the pressure treatment in the predictor step and of a selection of RK temporal schemes on the kinetic energy dissipation in Taylor-Green vortex flow. Next, the order of accuracy of a selection of temporal schemes is determined by execution of a self-convergence study for the lid-driven cavity.

\subsection{Error sources}
\label{sec:generalerrors}
The following possible error sources can be distinguished when using the RKSymmFoam numerical method: spatial and temporal discretisation errors,  the pressure error introduced by the application of a compact stencil in the pressure Poisson equation, the splitting error, the linearisation error of the convection term when implicit time integration is used, and other possible iteration errors.

Concerning the spatial discretisation, errors introduced by  mesh non-orthogonality and mesh skewness are zero in the presented cases with orthogonal grids. The discretisation of the convective and pressure terms may introduce errors in kinetic energy conservation \citep{feltenandlund}. As explained in section~\ref{sec:operator_requirements}, we use a skew-symmetric discretisation for the convective operator $\myC$ with interpolation weights of $\frac{1}{2}$ in order to conserve kinetic energy. This treatment is consistent with the study of \citet{feltenandlund} on kinetic energy conservation issues on collocated grids.

As explained in section~\ref{sec:pvcoupling}, we use the approximation $\Gamma_{c \rightarrow s} \Gamma_{s \rightarrow c} = \myI$ in order to obtain a compact Laplacian in the pressure Poisson equation. Due to the introduction of the compact Laplacian, a kinetic energy dissipation error is introduced, since the condition in Eq.~\ref{eq:OmegacGcMGammacs} is not exactly fulfilled.
This error is of $O(\Delta t \Delta h^2)$  when the compact Laplacian is based on the total pressure $\myp_c$ \citep{feltenandlund}. This approach is used in e.g. the classical projection method of \citet{chorin1968} or the standard PISO approach \citep{Issa1982}. The pressure error is reduced to $O(\Delta t^2 \Delta h^2)$  when the compact Laplacian is based on the pressure correction $\myp'_c$. This pressure correction approach is applied in \citet{Kan1986} type projection methods \citep{feltenandlund}, or the symmetry-preserving PISO method proposed in this paper.

For viscous flows, the discretisation of the diffusion operator $\myD$ plays a role in the prediction of the correct level of kinetic energy dissipation. As explained in section~\ref{sec:operator_requirements}, the diffusive operator $\myD$ which we use is symmetric positive-definite. Therefore, the diffusion term is strictly dissipative. The correct amount of kinetic energy is dissipated when the mesh resolution is sufficiently large for obtaining a fully resolved flow field. However, too little kinetic energy will be dissipated in case of an under-resolved flow  field. This will be further discussed in section~\ref{sec:energydissipation}.

Since we use five inner corrector steps in the present analyses, a negligible splitting error is obtained for the test cases considered in this section. For implicit time integration, a linearisation error in the computation of the convective face fluxes exists in principle. For the test cases in this section, we have used one, three, and five outer iterations. We found that the effect of the number of outer iterations on the results can be practically neglected. Therefore, we present the results for one outer iteration only. Other possible iteration errors (linear solvers) are negligible because of the application of very strict convergence criteria. As a result, it can be concluded that the following three error sources are present for the prediction of the correct level of kinetic energy in the subsequent analysis with orthogonal grids: 1) temporal discretisation errors,  2) the pressure error due to $\Gamma_{c \rightarrow s} \Gamma_{s \rightarrow c} = \myI$, and 3) the spatial discretisation error of the diffusion operator $\myD$.

\subsection{Taylor-Green vortex and lid-driven cavity}
The two-dimensional Taylor-Green vortex is frequently used in order to determine the  conservation properties of numerical schemes for incompressible flows. Taylor-Green vortex flow consists of an array of periodic vortices. The analytical solution of this flow is described by
\begin{equation}
    \label{eq:TG1}
    u(x,y,t) = \text{sin}(x) \, \text{cos}(y) \, e^{-2\nu t},
\end{equation}
\begin{equation}
    \label{eq:TG2}
    v(x,y,t) = -\text{cos}(x)\, \text{sin}(y)\, e^{-2\nu t},
\end{equation}
\begin{equation}
    \label{eq:TG3}
    p(x,y,t) = \frac{1}{4}(\text{cos}(2x) + \text{cos}(2y)) \, e^{-4\nu t},
\end{equation}
on a domain of $0 \leq x \leq 2\pi, \ 0 \leq y \leq 2\pi$ and periodic boundary conditions on the domain boundaries. The Reynolds number Re equals $1/\nu$. The Taylor-Green vortex flow is steady in the limit of inviscid flow with $\nu = 0$, or, $Re = \infty$. Consequently, the total kinetic energy $E_{kin}$ remains {\color{black} constant}, i.e., $\mathrm{d}E_{kin} / \mathrm{d}t = 0$. For viscous flow with $v > 0$,  the kinetic energy will decay with time, i.e., $\mathrm{d}E_{kin} / \mathrm{d}t < 0$. The decay rate can be calculated from the analytical solution.

The computations of the Taylor-Green vortex have been executed on a $64\times 64$ uniform orthogonal grid. The simulation time equals three integral time scales $\tau$, with $\tau  = 2\pi$. The maximum CFL-number equals $0.5$. We use five PISO corrections, which is much more than actually needed \citep{Issa1982}. However, this was done in order to reduce the splitting error to levels where it has no longer any significant contribution to the numerical dissipation rate.
Table \ref{tab:TGsettings} summarises the applied computational settings.

\begin{table}[t!]
    \centering
    \caption{Computational details for Taylor-Green vortex flow.}
    \label{tab:TGsettings}
    \begin{tabular}{ll}
        \hline
        Reynolds number Re  $= 1/\nu$ & $\infty$, 1000, 100 and 10 \\
        \hline
        Number of cells (x,y) & $64\times64$   \\
        Domain(x,y) & $2\pi,2\pi$ \\
        End time $t_{end}/ \tau$ & 3 \\
        Time step $\Delta t / \tau $ & 0.0015 \\
        Max CFL & 0.5 \\
        Spatial scheme convection & Midpoint \\
        Temporal scheme & Various schemes \\
        Inner (PISO) iterations & 5 \\
        Outer iterations & 1, 3 and 5 \\
        Initial field $t=0$ & Eqs. \ref{eq:TG1}, \ref{eq:TG2}, \ref{eq:TG3} \\
        \hline
    \end{tabular}
\end{table}



The lid-driven cavity (LDC) is also widely used in order to test numerical schemes. The two-dimensional LDC consists of a square cavity filled with an incompressible fluid with viscosity $\nu$. All four cavity walls have a length $L=1$. The upper wall is moving from left to right with a velocity $U_{lid}=1 $. As a result, a vortex is formed inside the cavity. We use a $64 \times 64$ and a $512 \times 512$  uniform Cartesian grid. The boundary conditions at the walls consist of a no-slip condition for velocity and
 a zero-gradient condition for pressure. The Reynolds number in the  lid-driven cavity equals Re $= 1/{\nu}$. The kinematic viscosity  $\nu$ equals 0.001. As a result, the Reynolds numbers equals 1000.
Table~\ref{tab:cavity} summarizes the computational details.
The time step $\Delta t$ has been incrementally reduced in the temporal convergence study.

\begin{table}[h!]
    \centering
    \caption{Computational details for the lid-driven cavity temporal consistency study.}
    \label{tab:cavity}
    \begin{tabular}{ll}
        \hline
        Reynolds number Re $={1}/\nu$ & 1000 \\
        \hline
        Number of cells (x,y) & $64\times 64$ and $512 \times 512$ \\
        Domain (x,y)          & 1,1 \\
        End time $t_{end}$    & 0.1 \\
        Time step $\Delta t$  & Various \\
        Max CFL               & 0.5 \\
        Spatial scheme convection          & Midpoint \\
        Temporal scheme       & Various \\
        Inner (PISO) iterations      & 5 \\
        Outer iterations      & 1, 3 and 5 \\
        Initial field $t=0$   & At rest \\
        \hline
    \end{tabular}
\end{table}

\subsection{Kinetic energy dissipation}
\label{sec:energydissipation}

\begin{figure}[h!]
    \centering
    \begin{subfigure}{0.496\linewidth}
        \centering
        \includegraphics{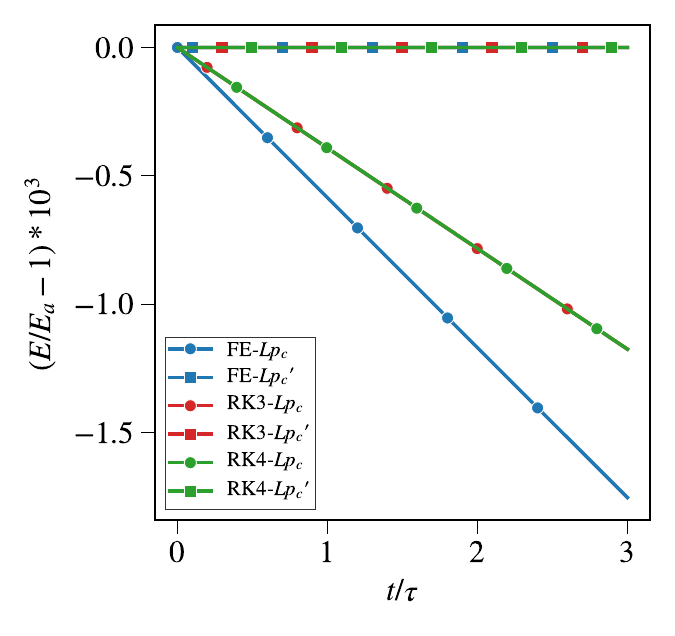}
        \caption{Explicit time integration}
        \label{fig:TG_inv_ex}
    \end{subfigure}
    \begin{subfigure}{0.496\linewidth}
        \centering
        \includegraphics{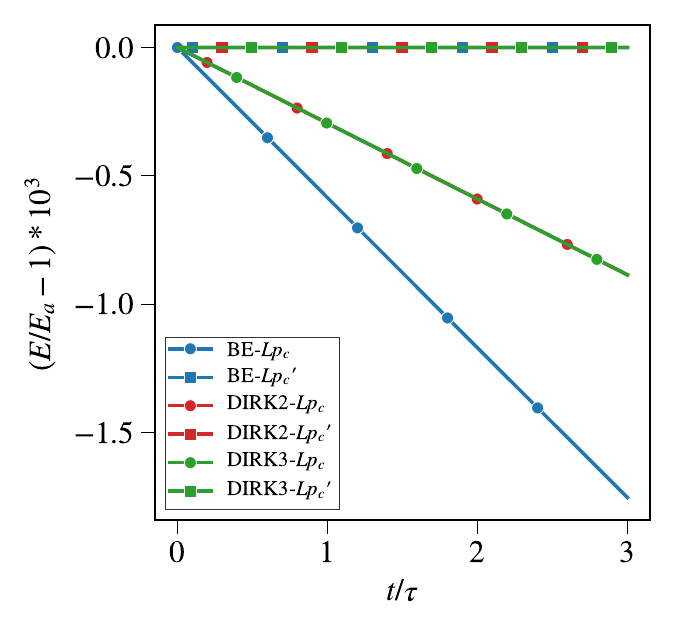}
        \caption{Implicit time integration}
        \label{fig:TG_inv_im}
    \end{subfigure}
    \caption{Scaled total kinetic energy error $(E-E_a)/E_a$ versus time for the inviscid Taylor-Green vortex.}
    \label{fig:TG_inv_ex_im}
\end{figure}

For the classical method of \citet{chorin1968} with $\myp_c^{p} = \mathbf{0}_{c,n}$
and  for a \citet{Kan1986} type projection method with  $\myp_c^{p} = \myp_c^{n}$, Fig.~\ref{fig:TG_inv_ex} presents the scaled total kinetic energy error $(E-E_a)/E_a$ versus time for the inviscid Taylor-Green vortex for three explicit time integration schemes. {\color{black}Here, the discrete analytical kinetic energy $E_a = E_0 e^{- 4 \nu t}$, where $E_0$ is the total discrete kinetic energy of the point-wise evaluated analytic velocity field at the start of the computation. E is the calculated total discrete kinetic energy of the simulation.}
The method of \citet{chorin1968} is indicated with $\myL \myp_c$ whereas the  \citet{Kan1986} type method is indicated with $\myL \myp'_c$.
Although we use a symmetry-preserving discretisation method, the results for the \citet{chorin1968} method still shows some dissipation, where the first order forward-Euler scheme shows somewhat more dissipation than the higher-order Runge-Kutta schemes.
As already explained, the approximation $\Gamma_{c \rightarrow s} \Gamma_{s \rightarrow c} = \myI$ was introduced in order to obtain a compact Laplacian in the pressure Poisson equation. As a consequence, a pressure error of $O(\Delta t \Delta h^2)$ is introduced for the \citet{chorin1968} method. When it is assumed that the difference between the first order forward Euler scheme and the $3^{rd}$ and $4^{th}$ order RK schemes provides an indication of the temporal errors, it is plausible to assume that the pressure error has the largest contribution to the dissipation in the considered results. Consequently, the $3^{rd}$ and $4^{th}$ order RK schemes provide identical results.
For the \citet{Kan1986} type method, the pressure error is reduced to $O(\Delta t^2 \Delta h^2)$. As a result, this method is practically dissipation free for all three applied time integration schemes.

Figure~\ref{fig:TG_inv_im} presents the corresponding results for three implicit time integration schemes. From a comparison of Figs.~\ref{fig:TG_inv_ex} and ~\ref{fig:TG_inv_im}, it can be concluded that the trends for the selected explicit and implicit schemes are very similar. It can be concluded that our symmetry-preserving PISO approach provides practically dissipation free results for all the applied temporal schemes when the pressure Poisson equation is based on a  pressure correction ($\myL \myp'_c$). Consequently, there is no artificial decay of the Taylor-Green vortex such that it remains steady in time. This explains why the backward Euler scheme is able to provide the same results as the selected DIRK schemes.

We have also performed computations with $\myp_c^{p} = 2 \myp_c^{n} - \myp_c^{n-1}$ in the predictor step, which should result in a pressure error of $O(\Delta t^3 \Delta h^2)$. However, checkerboarding occurs for this setting. Application of $\myp_c^{p} =  \myp_c^{n}$ in the predictor step yields a pressure error of $O(\Delta t^2 \Delta h^2)$. This setting provides results which are practically dissipation free. Therefore, we did not attempt to reduce this pressure error of $O(\Delta t^2 \Delta h^2)$ any further.

\begin{figure}[H]
    \centering
    \begin{subfigure}{0.496\linewidth}
        \centering
        \includegraphics{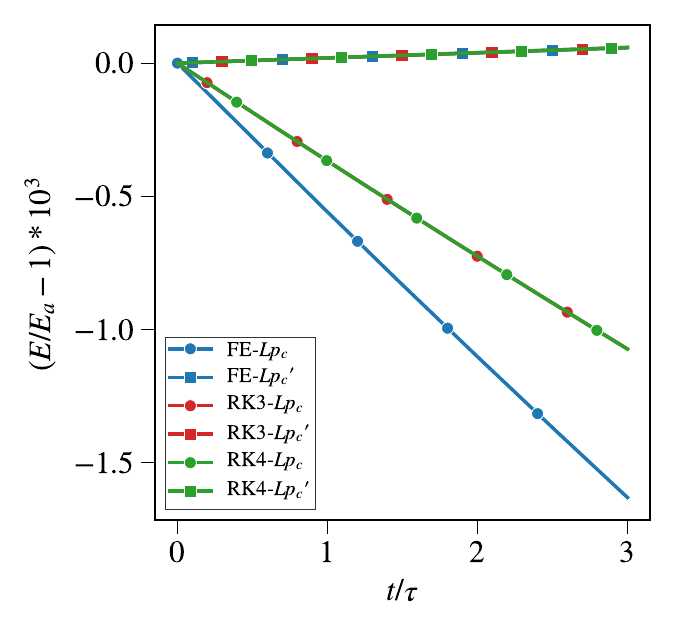}
        \caption{Explicit time integration}
        \label{fig:TG_inv}
    \end{subfigure}
    \begin{subfigure}{0.496\linewidth}
        \centering
        \includegraphics{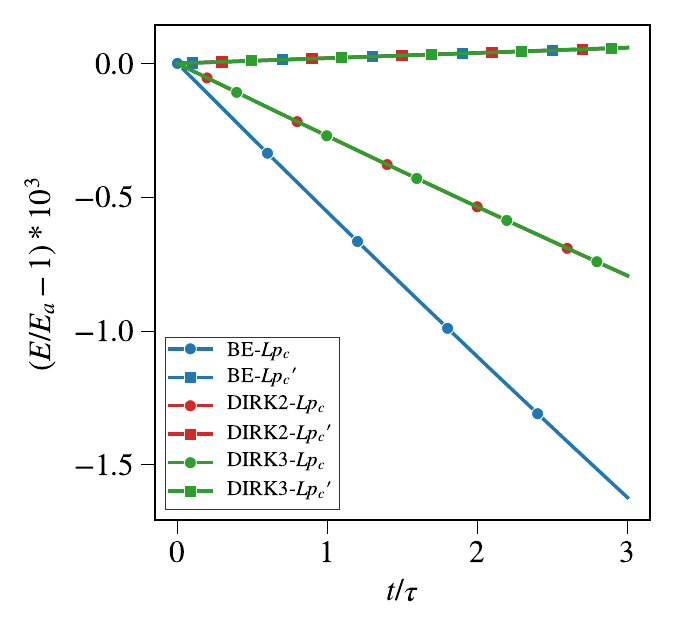}
        \caption{Implicit time integration}
        \label{fig:TG_visc}
    \end{subfigure}
    \caption{Scaled total kinetic energy error $(E-E_a)/E_a$ versus time for the Taylor-Green vortex at $Re = 1000$.}
    \label{fig:TG_visc_1000}
\end{figure}

We have also performed computations for the viscous Taylor-Green vortex at Re $= 1000$, see Fig.~\ref{fig:TG_visc_1000}. Here, $E_a$ now represents the corresponding viscous analytical solution.
For both the selected explicit and implicit temporal schemes, it can be observed that the dissipation rate is slightly underestimated when the $\myL \myp'_c$-based method is used.
From the corresponding results for the inviscid Taylor-Green vortex, we have concluded that this $\myL \myp'_c$-based method is practically free of numerical dissipation. Therefore, it can be concluded that the observed slight underestimation of the dissipation rate entirely corresponds to an underestimation of the physical energy dissipation as predicted by the diffusion operator $\myD$. This conclusion has been confirmed by a corresponding computation with higher mesh resolution. That is, for this higher mesh resolution, the correct amount of physical dissipation of kinetic energy is predicted (results not presented here).

Fig.~\ref{fig:TG_visc_1000} shows that the kinetic energy dissipation rate is overestimated
when the $\myL \myp_c$-based method is used. The results obtained with the first order Euler scheme show a somewhat larger overestimation than the RK and DIRK schemes. Since the underestimation of the physical dissipation by the diffusion operator $\myD$ is relatively negligible, the observed overestimation corresponds largely to numerical dissipation introduced by the pressure error and the temporal discretisation errors. Therefore, it can be concluded that error sources with an opposite sign are present when the $\myL \myp_c$-based method is used for the present case. Namely, the {\it numerical} dissipation caused by the pressure error and the temporal discretisation errors tends to overestimate the energy dissipation. In contrast, the diffusion operator $\myD$ underpredicts the {\it physical} dissipation rate when the mesh resolution is not sufficiently large.
We investigate the consequences of this aspect further in the next paragraph. {\color{black} Before doing so, we first present Table~\ref{tab:TG_visc_1000} which summarizes  the scaled total kinetic energy error $(E-E_a)/E_a$ at $t/\tau = 3$ for the Taylor-Green vortex at $Re = 1000$ for 1, 3 and 5 outer iterations. This table confirms that the effect of the number of outer iterations on the results can be neglected.
} 

\begin{table}[h!]
    \centering
    \caption{Scaled total kinetic energy error $(E-E_a)/E_a \times 10^3$ at $t/\tau = 3$ for the Taylor-Green vortex at $Re = 1000$.}
    \label{tab:TG_visc_1000}
    \begin{tabular}{lll}
        \hline
        nr. of outer iterations & $\myL \myp_c$ &  $\myL \myp'_c$ \\
        \hline
        1  &   -0.792883     & 0.058695     \\
        3  &   -0.793089     & 0.058692     \\
        5  &   -0.793089     & 0.058692     \\
        \hline
    \end{tabular}
\end{table}

\begin{figure}[H]
    \centering
    \begin{subfigure}{0.496\linewidth}
        \centering
        \includegraphics{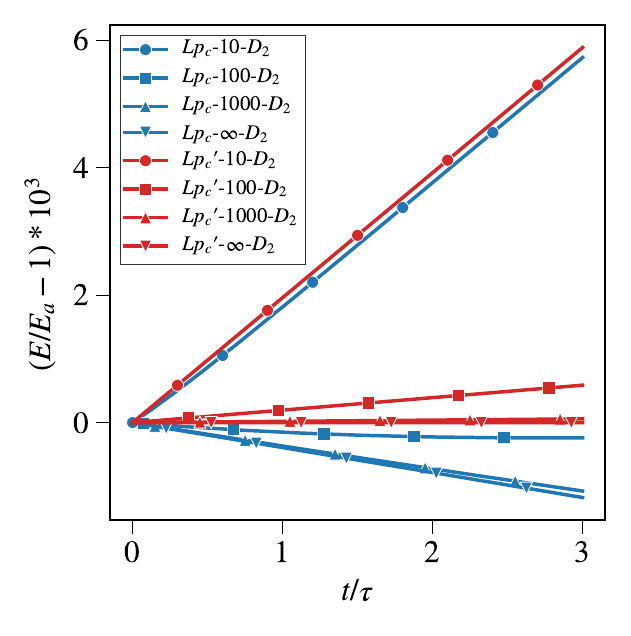}
        \caption{$\myL \myp_c$ and $\myL \myp'_c$ for $\myD2$ }
        \label{fig:TG_inv_variousRe1}
    \end{subfigure}
    \begin{subfigure}{0.496\linewidth}
        \centering
        \includegraphics{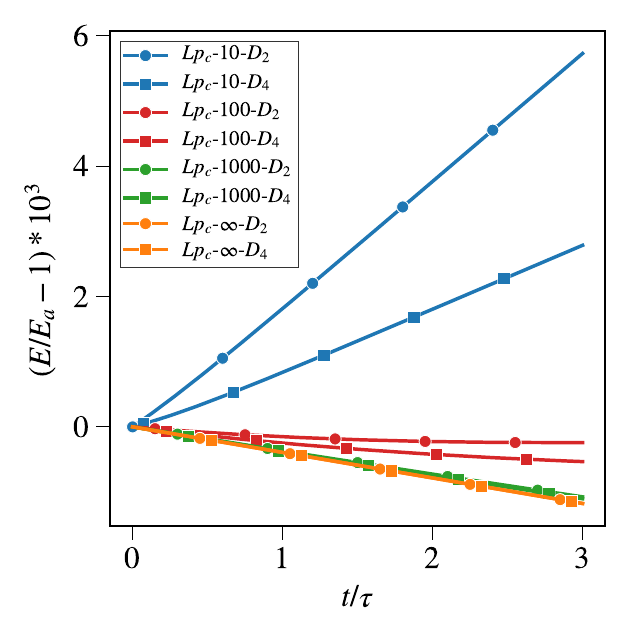}
        \caption{$\myD2$ and $\myD4$ for $\myL \myp_c$}
        \label{fig:TG_inv_variousRe2}
    \end{subfigure}
    \begin{subfigure}{0.496\linewidth}
        \centering
        \includegraphics{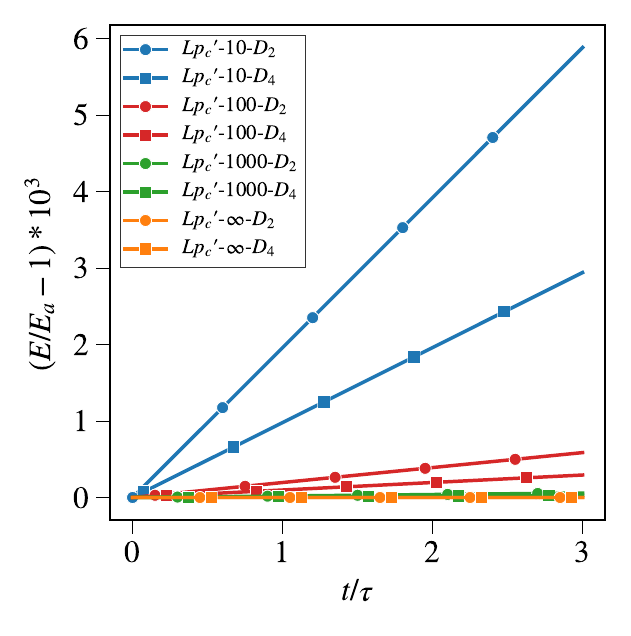}
         \caption{$\myD2$ and $\myD4$ for $\myL \myp'_c$}
        \label{fig:TG_inv_variousRe3}
    \end{subfigure}
      \caption{Scaled total kinetic energy error $(E-E_a)/E_a$ versus time for the Taylor-Green vortex for a range of Reynolds numbers. The $2^{nd}$ and $4^{th}$-order discretisaton for the diffusion operator $\myD$ are indicated with D2 and D4 respectively.}
    \label{fig:TG_inv_variousRe}
\end{figure}

For the Taylor-Green vortex, Fig.~\ref{fig:TG_inv_variousRe1} presents the scaled total kinetic energy error $(E-E_a)/E_a$ versus time for a range of Reynolds numbers. The results in this figure are obtained with the RK3 time integration scheme. We first consider the cases where the pressure Poisson equation is based on the pressure correction $\myp'_c$. We have already concluded that our method is effectively energy conserving when the pressure correction $\myp'_c$ is used. This is confirmed by the results in Fig.~\ref{fig:TG_inv_variousRe1} for the inviscid limit ($\text{Re} = \infty$). However, for the selected mesh resolution, the physical dissipation of kinetic energy is progressively underestimated by the discrete diffusion operator $\myD$ when the Reynolds number is successively reduced. 

Next, we discuss the results in Fig.~\ref{fig:TG_inv_variousRe1} obtained with a pressure Poisson equation which is based on the total pressure $\myp_c$. For the inviscid limit, the observed numerical dissipation of kinetic energy is largely caused by the presence of a pressure error of $O(\Delta t \Delta h^2)$. For the viscous cases,
the following competing effects are present: 1) overestimation of the energy dissipation as mainly caused by this pressure error, and 2) underprediction of energy dissipation by the diffusion operator $\myD$. This competition depends on the Reynolds number. Namely, the numerical dissipation is progressively compensated by an underestimation of the physical dissipation by the diffusion operator $\myD$ when the Reynolds number is successively reduced. As a result, by this compensation of errors with an opposite sign, the exact kinetic energy level will be predicted by the $\myL \myp_c$-based approach at a Reynolds number between 10 and 100 for the selected mesh resolution. In contrast, the energy preserving $\myL \myp'_c$-based approach will show an underprediction of kinetic energy dissipation for this Reynolds number. Although the $\myL \myp_c$-based approach shows a better prediction than the $\myL \myp'_c$-based for this specific situation, we consider this better prediction resulting from a compensation of errors of opposite sign as a generally undesirable situation.

The observation that the discrete diffusion operator $\myD$ results in an overprediction of the kinetic energy for low Reynolds numbers for the applied mesh resolution can be resolved by increasing the mesh resolution. However, since only the operator $\myD$ appears to cause a problem for low Reynolds numbers, an alternative solution to obtain better predictions is the application of higher order discretisation for $\myD$ only. Therefore, we investigate whether higher order discretisation for the diffusive operator $\myD$ is an effective measure to improve the predictive capabilities at low Reynolds numbers. More specifically, we test the $4^{th}$-order scheme presented in \ref{sec:AppHOD}.

A comparison of the application of $2^{nd}$ and $4^{th}$-order discretisaton of the diffusion operator $\myD$ is presented in  Fig.~\ref{fig:TG_inv_variousRe2} for the $\myL \myp_c$-based approach and in  Fig.~\ref{fig:TG_inv_variousRe3} for the $\myL \myp'_c$-based method. From these figures, it can be concluded that the $4^{th}$-order discretisaton results in a substantial improvement in the prediction of the kinetic energy for
$\text{Re} = 10$, and that  the $\myL \myp_c$ and $\myL \myp'_c$ methods provide practically the same prediction. For $\text{Re} = 100$, the $4^{th}$-order discretisaton causes a deterioration of the prediction for the $\myL \myp_c$-based approach, which can be understood by a smaller compensation of errors of opposite sign effect for this method. In contrast, the $4^{th}$-order discretisaton provides an improvement for the $\myL \myp'_c$-based method for $\text{Re} = 100$. For higher Reynolds numbers, the effect of the application of $4^{th}$-order discretisaton of $\myD$ is negligible. For these higher Reynolds numbers, the $\myL \myp'_c$-based approach provides the correct kinetic energy levels, whereas the $\myL \myp_c$-based method yields an underprediction of the kinetic energy. Overall, it can be concluded that the $4^{th}$-order discretisaton of $\myD$ provides improved predictions for low Reynolds numbers when the preferred $\myL \myp'_c$-based method is used.

\subsection{Temporal consistency}
\label{sec:temporalconsistency}

For both the $\myL \myp_c$ and $\myL \myp'_c$-based approaches, we perform $\Delta t$ self-convergence analyses for lid-driven cavity flow in order to determine the order of accuracy of the implemented numerical methods.
In these self-convergence analyses, we determine the temporal error as the difference between the temporally exact solution at very small time step and the corresponding solutions at larger time steps.  This procedure leads to an elimination of spatial errors which are independent of $\Delta t$. Note that the pressure error is both a spatial and temporal error, and will thus not vanish in the self-convergence analyses. Therefore, the $\Delta t$ self-convergence study exposes both the truncation error of the temporal scheme, as well as the pressure error.

\begin{figure}[t]
    \centering
    \begin{subfigure}{0.496\linewidth}
        \centering
        \includegraphics{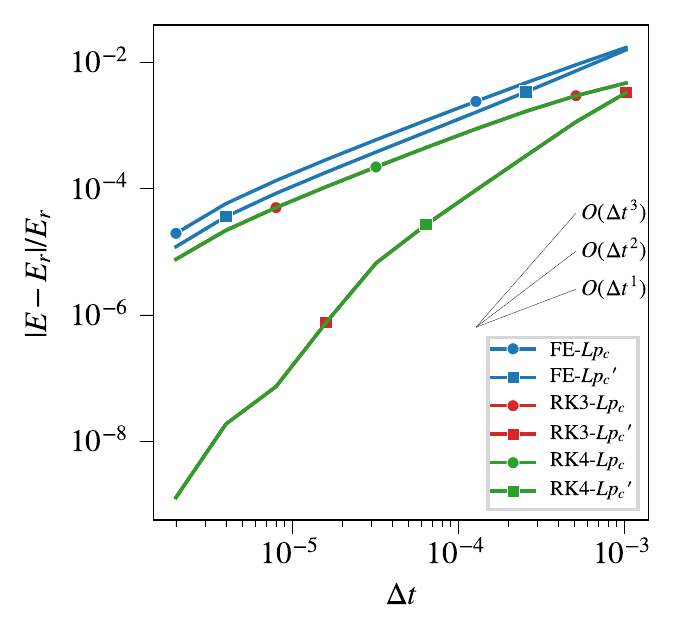}
        \caption{Explicit time integration schemes}
        \label{fig:LDC_explicit64}
    \end{subfigure}
    \begin{subfigure}{0.496\linewidth}
        \centering
        \includegraphics{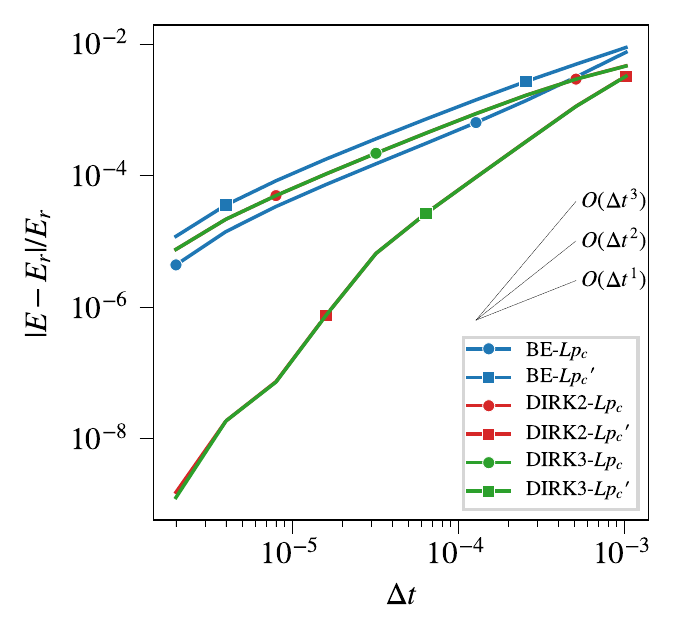}
        \caption{Implicit time integration schemes}
        \label{fig:LDC_implicit64}
    \end{subfigure}
    \caption{Scaled temporal error in the total kinetic energy as a function of the time step $\Delta t$ for the lid-driven cavity test case at  Re$=1000$ and 64x64 grid resolution. $E_r$ is the temporally exact solution at very small time step. }
    \label{fig:LDC_explicitandimplicit64}
\end{figure}

Fig.~\ref{fig:LDC_explicitandimplicit64} presents the scaled temporal error in the total kinetic energy as a function of the time step for the lid-driven cavity at  Re$=1000$, computed on a 64x64 Cartesian grid.
As demonstrated in \citet{komen2020}, the pressure error dominates over the temporal discretisation error on the applied 64x64 grid resolution. The pressure error is of $O(\Delta t  \Delta h^2)$ for the $\myL \myp_c$-based cases and of $O(\Delta t^2  \Delta h^2)$ for the $\myL \myp'_c$-based cases \citep{feltenandlund}. From Fig.~\ref{fig:LDC_explicitandimplicit64}, it can be concluded that $O(\Delta t)$ convergence is indeed observed for all selected explicit and implicit temporal schemes when the $\myL \myp_c$-based approach is used. That is, higher order temporal schemes suffer from a reduction of the temporal order to approximately one.
In contrast, in line with theory, $O(\Delta t^2)$ convergence is effectively obtained when the selected RK and DIRK temporal schemes are used together with the $\myL \myp'_c$-based method. Therefore, it can be concluded that, from an accuracy point of view, the more expensive RK4 and DIRK3 schemes have no added value compared to respectively the RK3 and DIRK2 schemes.

\begin{figure}[t]
    \centering
    \begin{subfigure}{0.496\linewidth}
        \centering
        \includegraphics{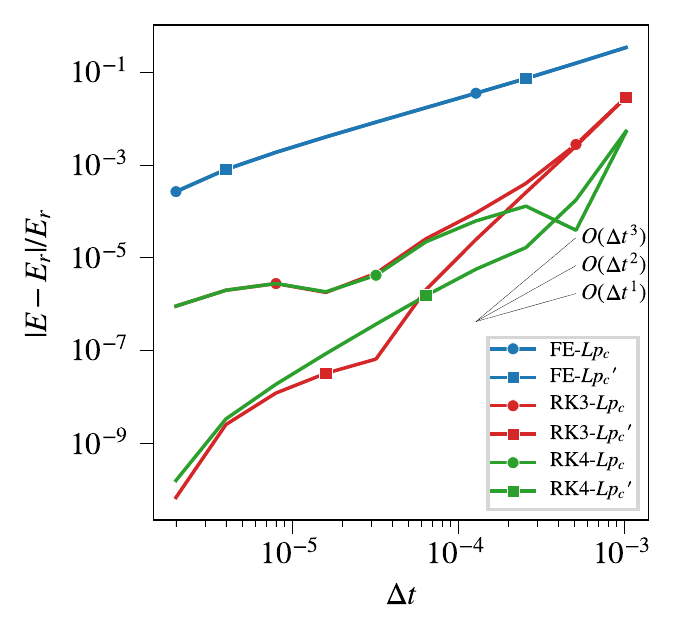}
        \caption{Explicit time integration schemes}
        \label{fig:LDC_explicit512}
    \end{subfigure}
    \begin{subfigure}{0.496\linewidth}
        \centering
        \includegraphics{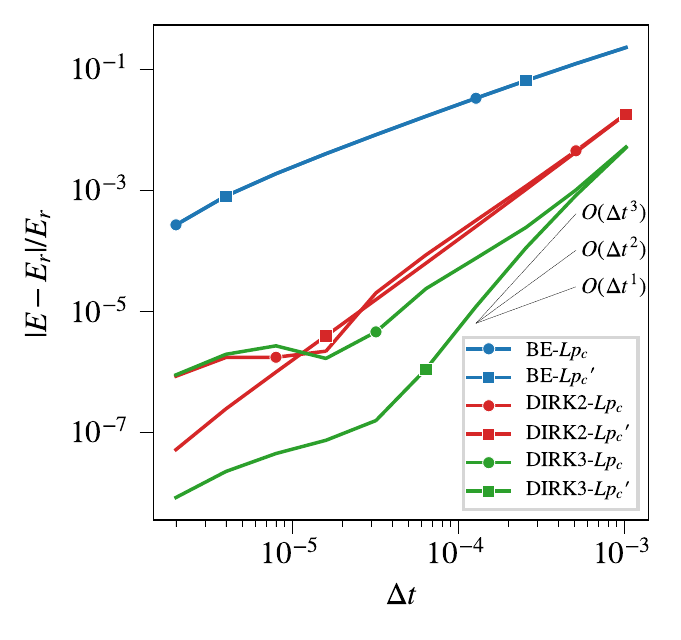}
        \caption{Implicit time integration schemes}
        \label{fig:LDC_implicit512}
    \end{subfigure}
    \caption{Scaled temporal error in the total kinetic energy as a function of the time step $\Delta t$ for the lid-driven cavity test case at  Re$=1000$ and 512x512 grid resolution. $E_r$ is the temporally exact solution at very small time step. }
    \label{fig:LDC_explicitandimplicit512}
\end{figure}

The pressure error is larger than the temporal discretisation error in the temporal convergence study using the 64x64 grid. The pressure error becomes negligible when a sufficiently refined grid is used. The theoretical order of accuracy of the applied temporal schemes should therefore be retrieved when the grid is sufficiently refined. Figure~\ref{fig:LDC_explicitandimplicit512} presents the results of a temporal convergence study using a mesh resolution of 512x512. As can be observed in this figure, the formal order of accuracy is obtained for the applied higher order temporal schemes for relatively large time steps. The pressure error becomes gradually larger than the temporal discretisation error when the time step is decreased. Consequently, the formal order of temporal accuracy of the applied higher order schemes reduces to approximately first order for the $\myL \myp_c$-based method and to approximately second order for the $\myL \myp'_c$-based method. This situation happens first for the RK4 and DIRK3 schemes.

\subsubsection{Effect of mesh non-uniformity, non-orthogonality, and skewness}
\label{sec:skewness}
\begin{figure}
   \centering
     \includegraphics[scale=0.2]{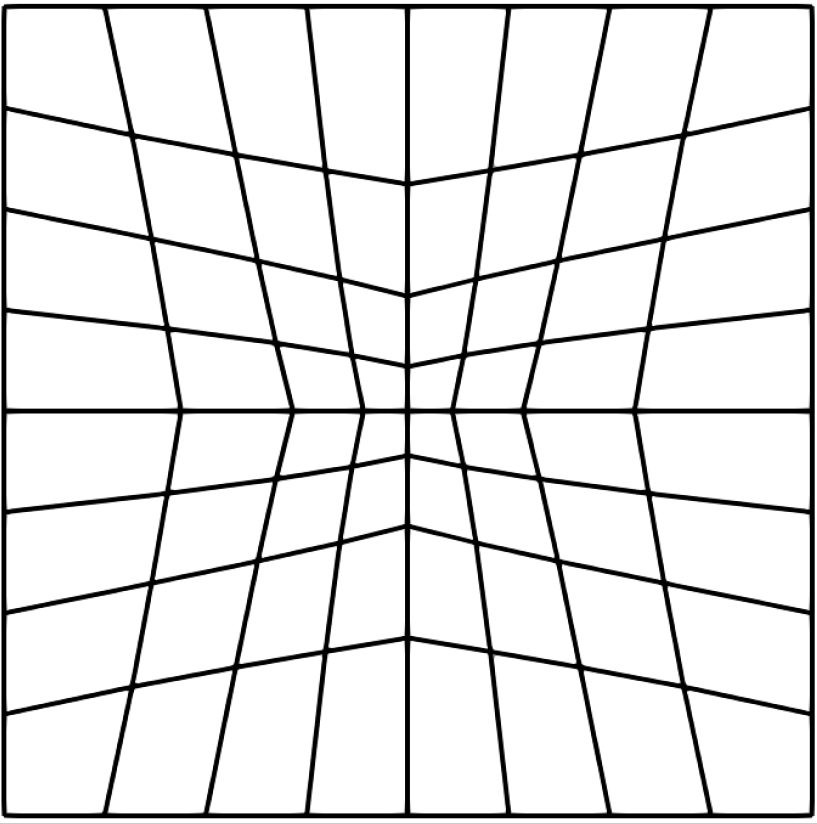}
    \caption{Mesh topology used to determine the effect from mesh non-uniformity, non-orthogonality, and skewness.}
    \label{fig:skewedmeshtopology}
\end{figure}

For modelling of flows in complex geometries, the unstructured collocated finite volume method is an attractive option. Unstructured polyhedral meshes and block-structured or unstructured hexahedral meshes are options for (q-)DNS or LES computations for flows in such complex geometries. \citet{komen2014} have shown that errors introduced by skewness and non-orthogonality can be avoided with good quality polyhedral meshes. Consequently, the accuracy obtained on good quality unstructured polyhedral meshes approaches the accuracy obtained on orthogonal hexahedral meshes closely. With the eventual aim of execution of (q-)DNS  or LES of flows in complex geometries in mind, we therefore also discuss results which we obtained using a distorted hexahedral grid. This distorted grid has been constructed in such a way that the grid includes both non-uniformity, non-orthogonality, as well as skewness. The applied grid topology is shown in Fig.~\ref{fig:skewedmeshtopology}. Like in the previously presented analyses, we use a mesh resolution of $64 \times 64$. The cell edge stretching in the vertical and horizontal mid plane has been chosen such that the ratio of the largest and smallest cell edge length equals four.

As mentioned in section~\ref{sec:generalerrors},
the following two kinetic energy conservation errors are present when uniform orthogonal grids are used:
errors resulting from 1) the temporal discretisation, and 2) the compact stencil in the pressure Poisson equation. When the distorted grid topology of Fig.~\ref{fig:skewedmeshtopology} is applied, the following two additional kinetic energy conservation errors are also introduced: errors caused by mesh non-orthogonality and mesh skewness. When a linear interpolation schemes is used, also a kinetic energy conservation error of  $O(f_x - \tfrac{1}{2}) \Delta h_f$ is present \citep{feltenandlund}, since the interpolation weight $f_x \neq \tfrac{1}{2}$ on non-uniform grids. However, since we use midpoint interpolation, this error source is not present in our analyses.
Due to the two additional error sources introduced by the distorted grid topology of Fig.~\ref{fig:skewedmeshtopology}, less favourable results are to be expected.

\begin{figure}[t]
    \centering
    \begin{subfigure}{0.496\linewidth}
        \centering
        \includegraphics{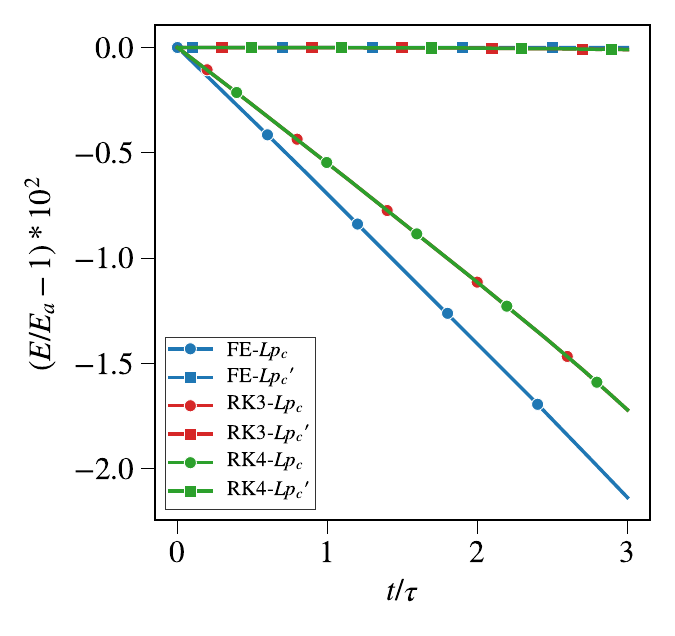}
        \caption{Explicit time integration}
        \label{fig:TG_inv_ex_skew}
    \end{subfigure}
    \begin{subfigure}{0.496\linewidth}
        \centering
        \includegraphics{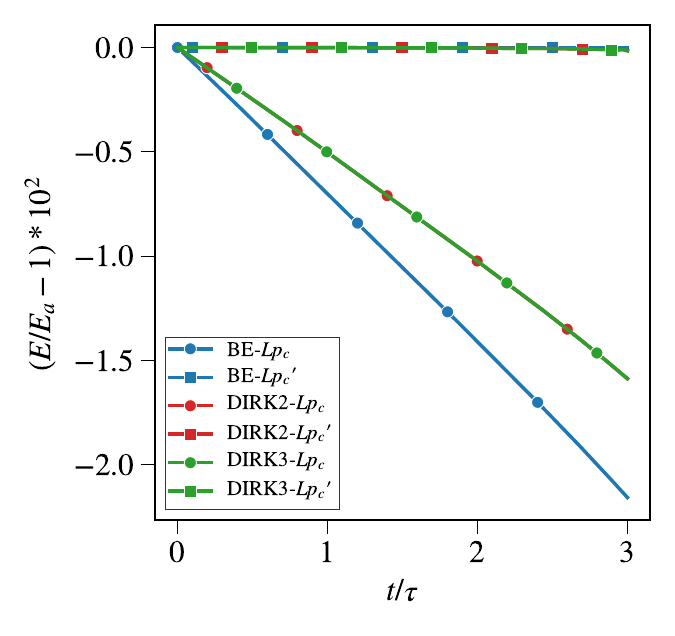}
        \caption{Implicit time integration}
        \label{fig:TG_inv_im_skew}
    \end{subfigure}
    \caption{Distorted hexahedral mesh: scaled total kinetic energy error $(E-E_a)/E_a$ versus time for the inviscid Taylor-Green vortex.}
    \label{fig:TG_inv_skew}
\end{figure}

For both the $\myL \myp_c$ and $\myL \myp'_c$-based methods, Fig.~\ref{fig:TG_inv_ex_skew} presents the scaled total kinetic energy error $(E-E_a)/E_a$  versus time for the inviscid Taylor-Green vortex for the three selected explicit time integration schemes. Similar as for the uniform Cartesian grid, the results obtained using the $\myL \myp_c$-based method show numerical dissipation, where the first order forward-Euler scheme shows more numerical dissipation than the higher-order Runge-Kutta schemes. When Fig.~\ref{fig:TG_inv_ex} and \ref{fig:TG_inv_ex_skew} are compared, it can be concluded that the numerical dissipation is an order of magnitude larger for the distorted hexahedral mesh when the $\myL \myp_c$-based method is used.
For the $\myL \myp'_c$-based method, for which pressure error is reduced to $O(\Delta t^2 \Delta h^2)$,
the obtained results are practically dissipation free for all three applied time integration schemes.
From a comparison of Figs.~\ref{fig:TG_inv_ex_skew} and \ref{fig:TG_inv_im_skew}, it can be observed that the results obtained with the selected explicit and implicit schemes are very similar. Overall, it can be concluded that, even for the distorted mesh of Fig.~\ref{fig:skewedmeshtopology}, our symmetry preserving PISO approach based on the application of a pressure correction ($\myL \myp'_c$) provides practically dissipation free results for all the applied temporal schemes for the inviscid Taylor-Green vortex.

\begin{figure}
    \centering
    \begin{subfigure}{0.496\linewidth}
        \centering
        \includegraphics{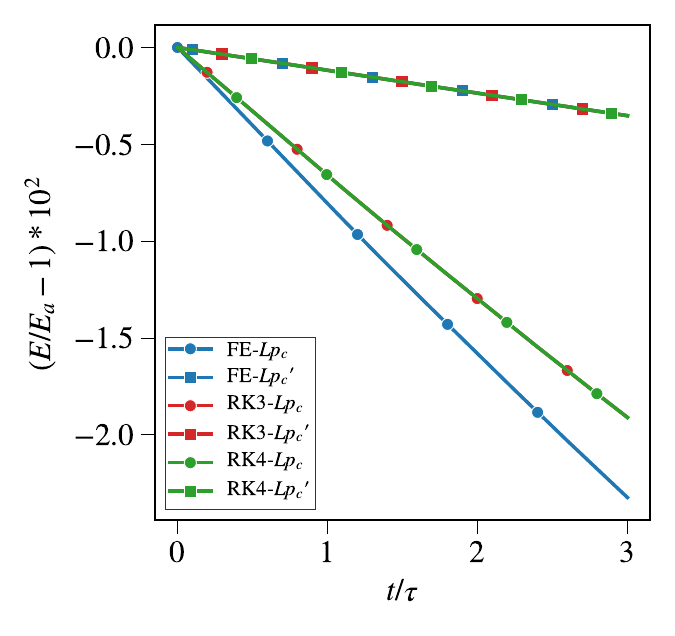}
        \caption{Explicit time integration}
        \label{fig:TG_visc_ex_skew}
    \end{subfigure}
    \begin{subfigure}{0.496\linewidth}
        \centering
        \includegraphics{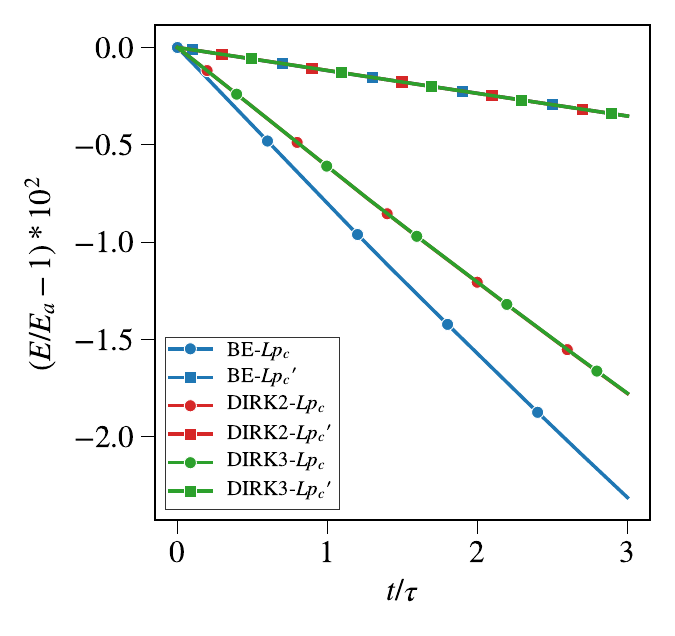}
        \caption{Implicit time integration}
        \label{fig:TG_visc_im_skew}
    \end{subfigure}
    \caption{Distorted hexahedral mesh: scaled total kinetic energy error $(E-E_a)/E_a$ versus time for the Taylor-Green vortex at $Re = 1000$.}
    \label{fig:TG_visc_1000_skew}
\end{figure}

The results obtained using the distorted hexahedral mesh for the viscous Taylor-Green vortex at Re $= 1000$ are shown in Fig.~\ref{fig:TG_visc_1000_skew}. For the corresponding results with the uniform Cartesian mesh (see Fig.~\ref{fig:TG_visc_1000}), we found  that the dissipation rate was slightly underestimated for the $\myL \myp'_c$-based method when using the selected explicit and implicit temporal schemes. This underestimation was caused by an underestimation of the  physical dissipation by the diffusion operator $\myD$ for the applied mesh resolution. In contrast, for the distorted mesh topology, the numerical dissipation introduced by the skewness and non-orthogonality errors results in a net overestimation of the dissipation rate for the $\myL \myp'_c$-based method. Due to the larger pressure error, the overestimation of the dissipation rate is even larger for the $\myL \myp_c$-based method, see Fig.~\ref{fig:TG_visc_1000_skew}. When Figs~\ref{fig:TG_visc_1000} and \ref{fig:TG_visc_1000_skew} are compared, it can be observed that the numerical dissipation rate is an order of magnitude larger for the distorted grid while using the $\myL \myp_c$-based method due to the presence of the skewness and non-orthogonality errors.

\begin{figure}
    \centering
    \begin{subfigure}{0.496\linewidth}
        \centering
        \includegraphics{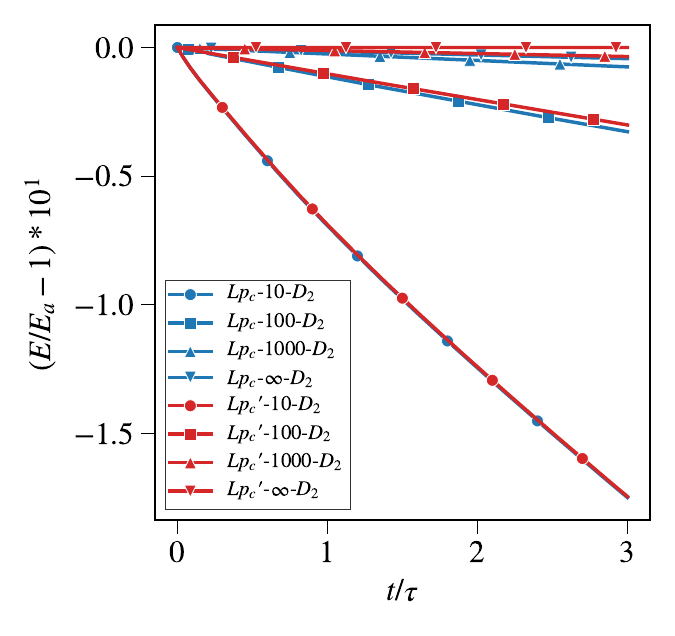}
        \caption{$\myL \myp_c$ and $\myL \myp'_c$ for $\myD2$ }
        \label{fig:TG_inv_variousRe1_skew}
    \end{subfigure}
    \begin{subfigure}{0.496\linewidth}
        \centering
        \includegraphics{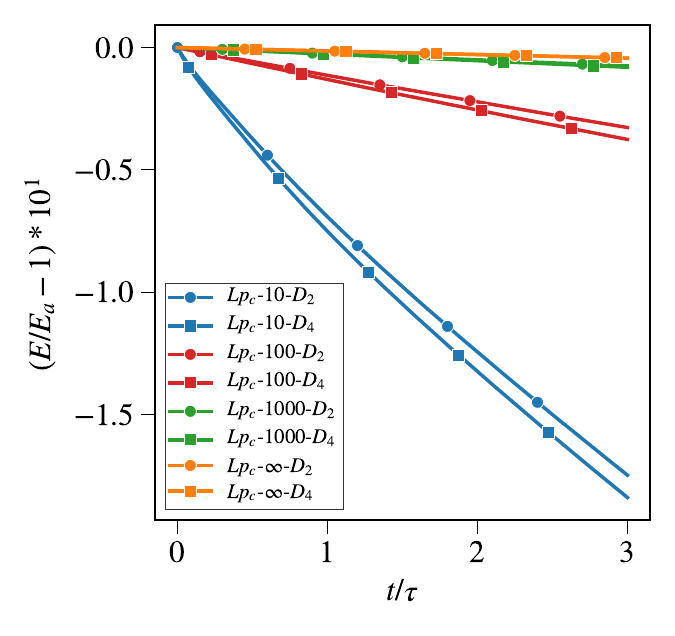}
        \caption{$\myD2$ and $\myD4$ for $\myL \myp_c$}
        \label{fig:TG_inv_variousRe2_skew}
    \end{subfigure}
    \begin{subfigure}{0.496\linewidth}
        \centering
        \includegraphics{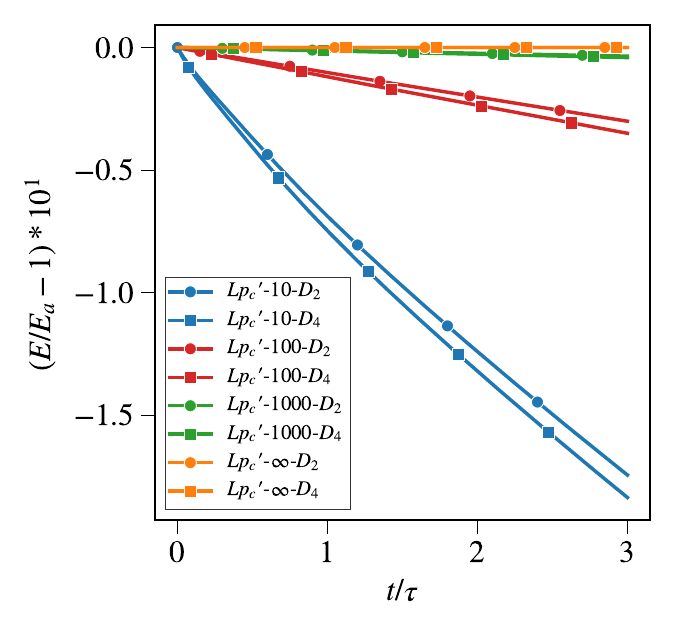}
         \caption{$\myD2$ and $\myD4$ for $\myL \myp'_c$}
        \label{fig:TG_inv_variousRe3_skew}
    \end{subfigure}
      \caption{Distorted hexahedral mesh and RK3 temporal scheme: scaled total kinetic energy error $(E-E_a)/E_a$ versus time for the Taylor-Green vortex for a range of Reynolds numbers. The $2^{nd}$ and $4^{th}$-order discretisaton for the diffusion operator $\myD$ are indicated with D2 and D4 respectively.}
    \label{fig:TG_inv_variousRe_skew}
\end{figure}

From a comparison of Fig.~\ref{fig:TG_inv_variousRe1} and  \ref{fig:TG_inv_variousRe1_skew}, it can be concluded that the numerical dissipation introduced by the skewness and non-orthogonality errors causes a transition from  an underestimation of the kinetic energy dissipation on the uniform Cartesian grid to a relatively large overestimation of the kinetic energy dissipation on the distorted hexahedral grid for $Re = 10$. This result is obtained for the $\myL \myp_c$-based as well as the $\myL \myp'_c$-based method. The considered effect is mainly a result of the skewness and non-orthogonality errors in the diffusion operator $\myD$. Namely, with increasing $Re$ number, the effect of the diffusion operator $\myD$ becomes smaller, and the obtained results converge gradually to the results obtained for the inviscid case with the RK3 temporal scheme on the same grid in Fig.~\ref{fig:TG_inv_skew}. As a result, our symmetry preserving discretisation based on $\myL \myp'_c$ provides practically dissipation free results for high $Re$ numbers.

Figures~\ref{fig:TG_inv_variousRe2_skew} and \ref{fig:TG_inv_variousRe3_skew} present a comparison of the usage of $2^{nd}$ and $4^{th}$-order discretisaton of the diffusion operator $\myD$ for respectively the $\myL \myp_c$-based  and $\myL \myp'_c$-based methods. For the uniform hexahedral grid, it was concluded
that the application of $4^{th}$-order discretisaton of $\myD$ provides improved predictions for our preferred $\myL \myp'_c$-based method for low Reynolds numbers (see Fig~\ref{fig:TG_inv_variousRe}). In contrast, the application of $4^{th}$-order discretisaton for $\myD$ yields somewhat worse predictions compared to the $2^{th}$-order discretisaton method for both the $\myL \myp_c$-based  and $\myL \myp'_c$-based methods for low $Re$ numbers when using the distorted hexahedral grid. This is caused by the fact that the local truncation error of the applied  $4^{th}$-order discretisation method reduces to first-order for irregular grids. For high Reynolds numbers, the contribution of the diffusion operator $\myD$ becomes less important, and the results converge gradually to the results obtained for the inviscid case with the RK3 temporal scheme on the same grid in Fig.~\ref{fig:TG_inv_skew}.

\begin{figure}
    \centering
    \begin{subfigure}{0.496\linewidth}
        \centering
        \includegraphics{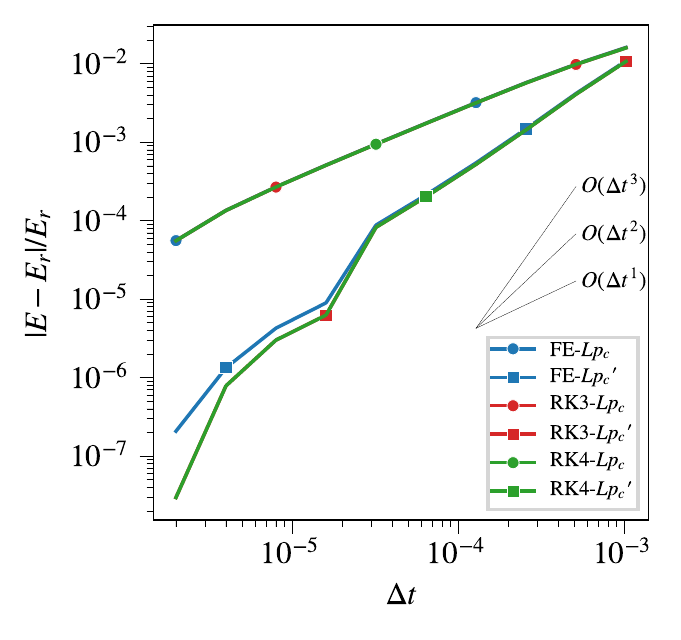}
        \caption{Explicit time integration schemes}
        \label{fig:LDC_explicit64_skew}
    \end{subfigure}
    \begin{subfigure}{0.496\linewidth}
        \centering
        \includegraphics{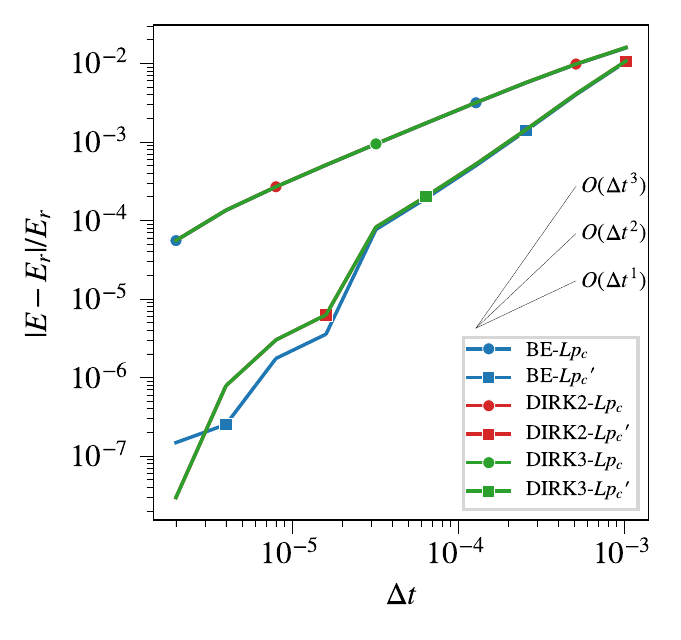}
        \caption{Implicit time integration schemes}
        \label{fig:LDC_implicit64_skew}
    \end{subfigure}
    \caption{Distorted hexahedral mesh: scaled temporal error in the total kinetic energy as a function of the time step $\Delta t$ for the lid-driven cavity test case at  Re$=1000$ and 64x64 grid resolution. $E_r$ is the temporally exact solution at very small time step. }
    \label{fig:LDC_explicitandimplicit64_skew}
\end{figure}

The scaled temporal error in the total kinetic energy as a function of the time step $\Delta t$ for the lid-driven cavity test case at  Re $ =1000$ is presented in Fig.~\ref{fig:LDC_explicitandimplicit64_skew} for the distorted $64 \times 64$ mesh. From the corresponding Cartesian mesh results
(Fig.~\ref{fig:LDC_explicitandimplicit64}), we concluded that $O(\Delta t)$ convergence is obtained for all selected explicit and implicit temporal schemes for
the $\myL \myp_c$-based approach due to the presence of a pressure error of $O(\Delta t \Delta h^2)$. When Figs~\ref{fig:LDC_explicitandimplicit64} and \ref{fig:LDC_explicitandimplicit64_skew} are compared, it can be concluded that quite similar temporal convergence behaviour is observed when using the $\myL \myp_c$-based approach and the distorted hexahedral grid. In contrast, slightly worse convergence behaviour is observed for the higher order RK and DIRK schemes when the $\myL \myp'_c$-based approach is used together with the distorted hexahedral mesh.

\subsubsection{Overall performance assessment}
From the above results, it can be concluded that our new symmetry-preserving PISO approach for implicit time stepping shows very similar predictive performance as the applied symmetry-preserving projection approach for explicit time stepping. The RK3 temporal scheme requires the solution of three pressure Poisson equations per time step. Solution of four pressure Poisson equations is needed per time step when the DIRK2 schemes is used, since two PISO iterations per stage are sufficient for second order temporal accuracy. The major part of the computer time is required for solving the pressure Poisson equations. Therefore, it can be expected that the considered PISO approach with DIRK2 requires approximately 33\% more computing time per time step than the projection approach with RK3. However, it may be noted that the selected iterative convergence criteria for the solution of the pressure Poisson equations can have a substantial effect on this number.

{\color{black} More complex, industrial, test cases are needed in order to explore the possible benefits of the presented symmetry-preserving PISO approach
in terms of stability, accuracy and efficiency. The analysis for such test cases is outside the scope of the present paper. Therefore, this is recommended for further research.}


\section{Summary and Conclusions}
\label{sec:conclusions}
We have implemented the symmetry-preserving discretisation method for unstructured collocated grids of \citet{trias2014} in the open source code \OF. This method is based on a projection method with explicit time integration. We have extended this approach with a new symmetry-preserving second-order time-accurate PISO-based method with implicit time integration. Furthermore, we have implemented a Butcher tableau in \OF, such that we have access to a large family of ERK and  DIRK schemes for time integration.
We have presented analyses for Taylor-Green vortex cases in order to study the energy conservation properties of our RKSymmFoam solver. Furthermore, we have presented temporal consistency analyses for a range of temporal schemes in RKSymmFoam for a lid-driven cavity test case. The presented analyses have been performed for both a uniform Cartesian mesh as well as a distorted hexahedral mesh.

{\it Conclusions for the analyses using the uniform Cartesian meshes}\\
Concerning the prediction of the correct level of kinetic energy in the presented Taylor-Green analysis using Cartesian grids, the following three error sources are introduced by the application of 1) the temporal discretisation, 2) the approximation $\Gamma_{c \rightarrow s} \Gamma_{s \rightarrow c} = \myI$, and 3) the discretisation of the diffusion operator $\myD$. The approximation $\Gamma_{c \rightarrow s} \Gamma_{s \rightarrow c} = \myI$ was introduced in order to obtain a compact Laplacian in the pressure Poisson equation. Consequently, a pressure error of $O(\Delta t \Delta h^2)$ is introduced when this Laplacian is based on the total pressure ($\myL \myp_c$-based method). This pressure error is of  $O(\Delta t^2 \Delta h^2)$ when the Laplacian is based on the pressure correction $p'_c$ ($\myL \myp'_c$-based method).

Although our numerical method is based on a symmetry-preserving discretisation method, we observed that the $\myL \myp_c$-based method still yields some numerical dissipation in the presented inviscid Taylor-Green vortex cases. This artificial dissipation is mainly caused by the pressure error. When the $\myL \myp'_c$-based method is used,  the pressure error is reduced to $O(\Delta t^2 \Delta h^2)$,  and the computational results are practically dissipation free. Because of its energy conservation properties, the $\myL \myp'_c$-based approach is the preferred option. More specifically, we concluded that our new symmetry-preserving second-order time-accurate PISO-based method is practically free of numerical dissipation. {\color{black}
It has been observed that symmetry-preserving discretisation combined with a pressure correction in the pressure Poisson equation is needed for obtaining kinetic energy conservation on collocated grids.}

Subsequently, we have performed Taylor-Green vortex analyses where we reduced the Reynolds number stepwise from $Re = \infty$ to $Re= 10$. From these analyses, we concluded that, for the prediction of the correct level of kinetic energy, the mesh resolution requirements become more restrictive for low Reynolds numbers ($Re$ smaller than $O(100)$). Furthermore, these requirements are imposed by the diffusion term. The applied  $4^{th}$-order D4 discretisaton for the diffusion operator $\myD$ provided consistently improved predictions for the $\myL \myp'_c$-based method for low Reynolds numbers. In contrast, no consistent improvements were found for the $\myL \myp_c$-based method. This is caused by the presence of errors of opposite sign for this method for low Reynolds numbers. The D4 and D2 discretisation of the diffusion operator $\myD$ provide practically identical results for high Reynolds numbers.

Due to the presence of a pressure error of $O(\Delta t \Delta h^2)$ for the $\myL \myp_c$-based approach, all selected explicit and implicit higher order temporal schemes suffer from a reduction of the temporal order to approximately one when using the $\myL \myp_c$-based approach.
Since the magnitude of the pressure error is reduced to $O(\Delta t^2 \Delta h^2)$ for the $\myL \myp'_c$-based approach, second-order temporal convergence is effectively obtained when the selected higher order RK and DIRK temporal schemes are used. Consequently, from an accuracy point of view, the more expensive RK4 and DIRK3 schemes do not have added value compared to respectively the RK3 and DIRK2 schemes.

{\it Conclusions for the analyses using the distorted hexahedral mesh}\\
In addition to the kinetic energy conservation errors resulting from the temporal discretisation and the compact stencil in the pressure Poisson equation, additional errors sources are introduced due to mesh non-orthogonality and mesh skewness while using the applied distorted hexahedral mesh.
It was concluded that these two additional errors sources yield a substantial increase of the numerical energy dissipation in the Taylor-Green test cases at low Reynolds numbers.
When the Reynolds number was stepwise increased in these test cases, we found that our symmetry preserving discretisation method based on $\myL \myp'_c$ provides practically dissipation free results for high Reynolds numbers. Consequently,  the observed increased numerical dissipation is mainly caused by the skewness and non-orthogonality errors in the diffusion operator $\myD$.
From the performed temporal consistency analyses, we concluded that the additional errors source introduced by the distorted mesh further reduce the order of the applied higher order temporal schemes.

In an attempt to obtain a pressure error of $O(\Delta t^3 \Delta h^2)$, we have tested the application of  $\myp_c^{p} = 2 \myp_c^{n} - \myp_c^{n-1}$ in the predictor step. However, this resulted in checkerboarding, indicating the inherent limitations of the collocated grid arrangement.

As further research, we intend to perform turbulent channel flow analyses as a next step to more complex test cases for the further assessment of the implemented symmetry-preserving discretisation methods.

\section{Acknowledgement}
{\color{black} F.X. Trias is supported by the {\it Ministerio de Econom\'{i}a y Competitividad}, Spain (ENE2017-88697-R). In addition to that, }this research did not receive any specific grant from funding agencies in the public, commercial, or not-for-profit sectors.


\newpage
\appendix

\section{implemented D4 diffusion operator}
\label{sec:AppHOD}
This section presents the implementation of the D4 diffusion operator in {\OF}. We used this D4 operator in the present study in order to determine whether higher order discretisation of the diffusion operator is an effective measure for better prediction of the correct levels of kinetic energy in the considered Taylor-Green vortex test cases for low Reynolds numbers. This D4 scheme provides fourth order accuracy on uniform Cartesian meshes. However, like for the implemented second order D2 scheme, the accuracy of the D4 scheme reduces to approximately first order for irregular grids. In contrast to this D2 scheme, the implemented D4 scheme is not symmetric for distorted grids.

\begin{figure}[ht]
	\centering
	\includegraphics[width=0.9 \textwidth]{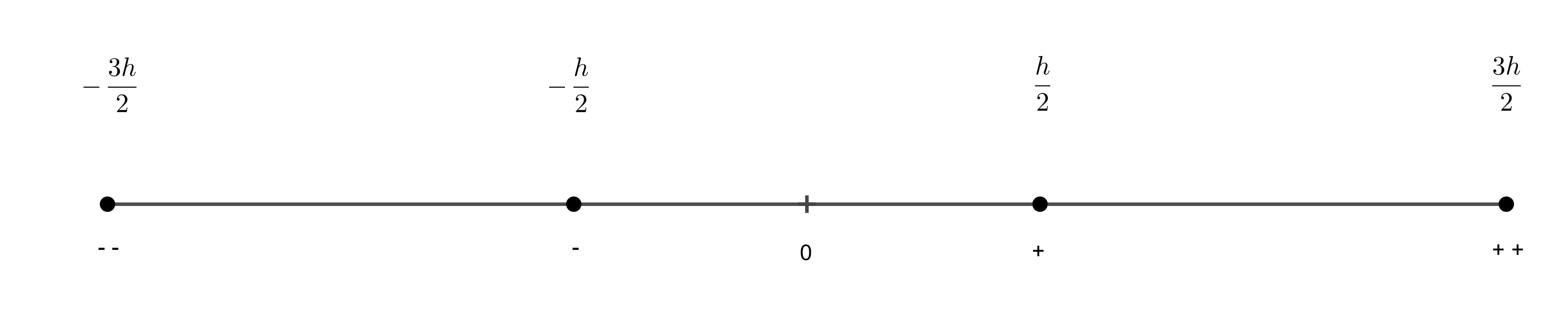}
	\caption{one-dimensional uniform grid}
	\label{fig:1dGrid}
\end{figure}

In order to obtain higher order discretisation of the diffusion term, we first consider the one-dimensional uniform grid of Fig.~\ref{fig:1dGrid}.

Using Taylor expansions for all cell centers around face 0, the following fourth order face gradient at face 0 can be obtained
\begin{equation}
\label{eq:gradJannes}
   \left[  \frac{du}{dx} \right]_0 = \frac{u_{--} - 27u_{-} + 27u_{+} - u_{++}}{24h} + O(h^4).
\end{equation}
For the implementation of this scheme in OpenFOAM, we use a combination of the cell-centered gradient $\myG_c$ and the face gradient $\myG_s$. In one dimension, the  cell-centered gradient $\myG_c$ at $+$ equals
\begin{equation}
    [\myG_c u_c]_+ = \frac{u_{++}-u_{-}}{2h},
\end{equation}
and similarly
\begin{equation}
    [\myG_c u_c]_- = \frac{u_{+}-u_{--}}{2h}.
\end{equation}
Using the midpoint interpolation operator $\Gamma_{c\rightarrow{s}}$, we obtain
\begin{equation}
\label{eq:IntGradU}
    [\Gamma_{c\rightarrow{s}}\myG_c u_c]_0 = \frac{-u_{--} - u_{-} + u_{+} + u_{++}}{4h}.
\end{equation}
The face gradient $\myG_s$ at $0$ is given by
\begin{equation}
\label{eq:FaceGradU}
    [\myG_s u_c]_0 = \frac{u_+-u_-}{h}.
\end{equation}
Therefore, the considered fourth order face gradient at face 0 can be implemented using
\begin{equation}
\label{eq:HOD}
 \left[  \frac{du}{dx} \right]_0 = \frac{u_{--} - 27u_{-} + 27u_{+} - u_{++}}{24h} + O(h^4) =
    \frac{7}{6}[\myG_s u_c]_0 + \frac{-1}{6}[\Gamma_{c\rightarrow{s}}\myG_c u_c]_0 + O(h^4).
\end{equation}

\begin{figure}[ht]
	\centering
	\includegraphics[width=0.6 \textwidth]{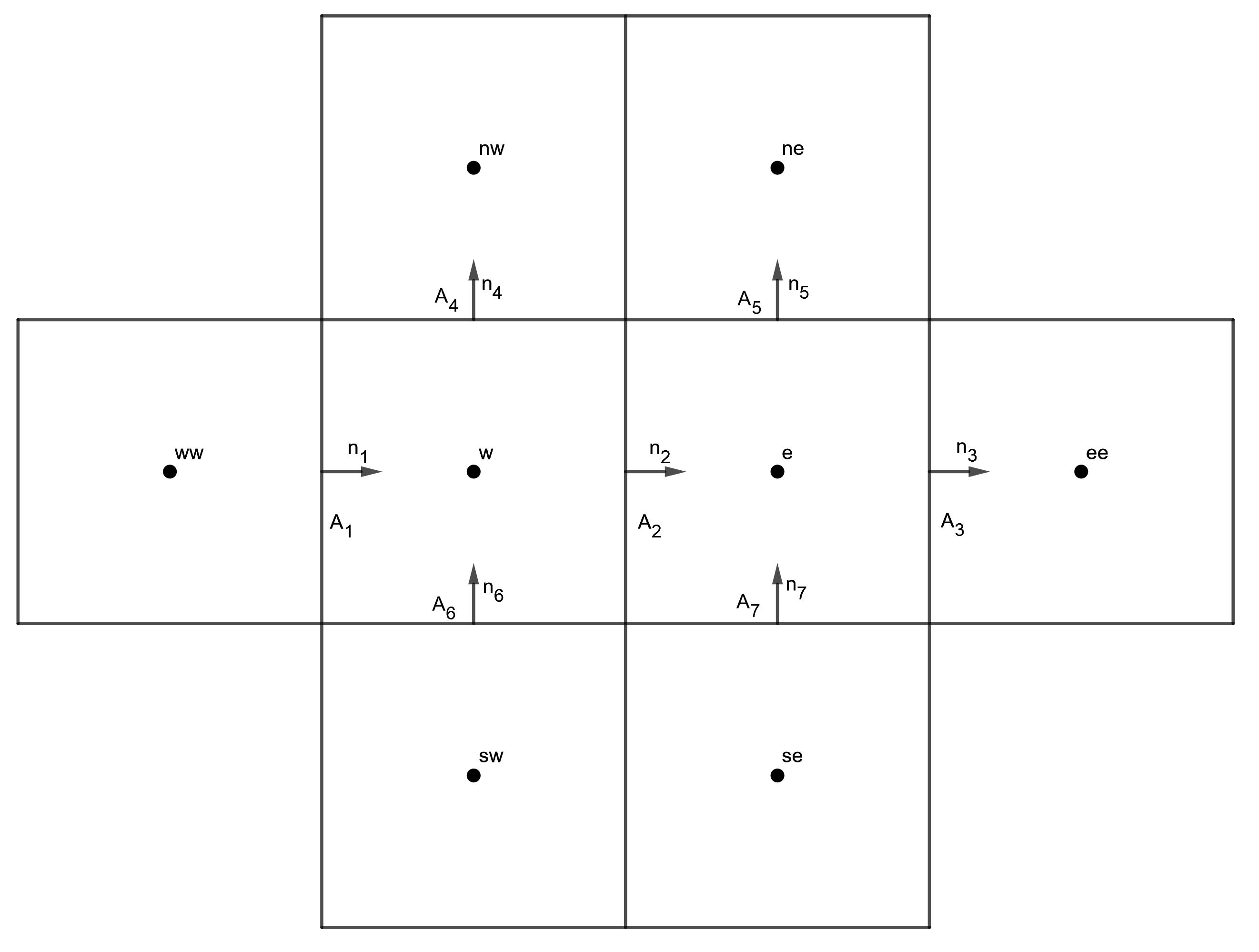}
	\caption{Two-dimensional uniform Cartesian grid for computing higher order gradient at face 2}
	\label{fig:2DGridGrad}
\end{figure}

Using the considered cell-centered gradient $\myG_c$ and the face gradient $\myG_s$, which are standard implemented operators in {\OF}, the considered fourth order gradient scheme can be generalised to three-dimensional unstructured grids as follows:
the cell-centered gradient $\myG_c$ which is implemented in {\OF} is a Gaussian gradient. That is, the cell-centered velocity gradient $[\myG_c \myu_c)]_i$ at cell $i$ is calculated as
\begin{equation}
    [\myG_c \myu_c]_i = \frac{1}{\Omega_{i,i}} \sum_{f\in F_f(i)}  \Big[\frac{\myu_i + \myu_j}{2}\Big]  \otimes \textbf{n}_f A_f.
\end{equation}
In order to demonstrate that equation \ref{eq:HOD} holds when this definition of $\myG_c$ is used, the two-dimensional uniform cartesian grid of Fig.~\ref{fig:2DGridGrad} is considered,
The cell-centered gradient $[\myG_c \myu_c]_w$ for the cell $w$ is computed as follows: of this cell, the term pertaining to face 4 is determined by
$$
[\myG_c \myu_c]_{w,4}  =
    \begin{pmatrix}
        \frac{u_{w}+u_{nw}}{2} \\
        \frac{v_{w}+v_{nw}}{2}
    \end{pmatrix}
    \begin{pmatrix}
        n_{4,x} && n_{4,y}
    \end{pmatrix}
    A_4
$$
  \begin{equation}
     = A_4
    \begin{pmatrix}
        0 && \frac{u_{w}+u_{nw}}{2} \\
        0 && \frac{v_{w}+v_{nw}}{2}
    \end{pmatrix}
\end{equation}
Analogously, the corresponding term for face 1 equals
\begin{equation}
    [\myG_c \myu_c]_{w,1} = -A_1
    \begin{pmatrix}
        \frac{u_{ww}+u_{w}}{2} && 0 \\
        \frac{v_{ww}+v_{w}}{2} && 0
    \end{pmatrix}
\end{equation}
The corresponding expressions for faces 2 and 6 are determined in a similar way. Subsequently,
the cell-centered gradient $[\myG_c \myu_c]_w$ is computed as the sum of the four face contributions  divided by the cell volume, that is,
\begin{equation}
    [\myG_c \myu_c]_w = \frac{1}{\Omega_{w,w}}
    \begin{pmatrix}
        -A_1 \frac{u_{ww}+u_{w}}{2} + A_2 \frac{u_{w}+u_{e}}{2} &&
        A_4 \frac{u_{w}+u_{nw}}{2} - A_6 \frac{u_{sw}+u_{w}}{2} \\
        -A_1 \frac{v_{ww}+v_{w}}{2} + A_2 \frac{v_{w}+v_{e}}{2} &&
        A_4 \frac{v_{w}+v_{nw}}{2} - A_6 \frac{v_{sw}+v_{w}}{2} \\
    \end{pmatrix}
\end{equation}
The cell-centered gradients of cells $w$ and $e$ are subsequently interpolated to face 2, that is,
\begin{equation}
\label{eq:InterpolGc}
\begin{split}
    [\Gamma_{c\rightarrow{s}}\myG_c \myu_c)]_2 & = \frac{[\myG_c \myu_c]_w}{2} \cdot n_2 + \frac{[\myG_c \myu_c]_e}{2} \cdot n_2 \\
    & =
    \begin{pmatrix}
        -\frac{A_1}{\Omega_{w,w}} \frac{u_{ww}+u_{w}}{4} + \frac{A_2}{\Omega_{w,w}} \frac{u_{w}+u_{e}}{4}
        -\frac{A_2}{\Omega_{e,e}} \frac{u_{w}+u_{e}}{4} + \frac{A_3}{\Omega_{e,e}} \frac{u_{e}+u_{ee}}{4} \\
        -\frac{A_1}{\Omega_{w,w}} \frac{v_{ww}+v_{w}}{4} + \frac{A_2}{\Omega_{w,w}} \frac{v_{w}+v_{e}}{4}
        -\frac{A_2}{\Omega_{e,e}} \frac{v_{w}+v_{e}}{4} + \frac{A_3}{\Omega_{e,e}} \frac{v_{e}+v_{ee}}{4}
    \end{pmatrix}
\end{split}
\end{equation}
For the considered uniform cartesian grid  $A_1=A_2=A_3=A_f$,  $\Omega_{e,e}=\Omega_{w,w}=A_f^2$ and $A_f = h$, with h the distance between  cell centers. Therefore, Eq~\ref{eq:InterpolGc} simplifies to
\begin{equation}
    [\Gamma_{c\rightarrow{s}}\myG_c \myu_c]_2 = \frac{-\myu_{ww} - \myu_{w} + \myu_{e} + \myu_{ee}}{4h},
\end{equation}
which corresponds equation \ref{eq:IntGradU}.

In {\OF}, the face gradient $[\myG_s \myu_c]_2$ at face 2 is calculated as
\begin{equation}
    [\myG_s \myv_c]_2 = \frac{\myv_e - \myv_w}{|\textbf{n}_2\cdotp \overrightarrow{we}|},
\end{equation}
which in this example simplifies to
\begin{equation}
    [G_s\myv_c)]_2 = \frac{\myv_e - \myv_w}{h},
\end{equation}
just like equation \ref{eq:FaceGradU}. Using the same weights factors as used in Eq.~\ref{eq:HOD}, the higher order discretisation of the diffusive term of Eq.~\ref{eq:HOD} is retrieved.

{\color{black}
For the one-dimensional uniform grid of Fig.~\ref{fig:1dGrid}, an analytical assessment of the D2 and D4 operators can be made as follows: application of the continuous Laplacian operator on the single wave fuction $\phi(x) = e^{ikx}$ yields $d^2 \phi / d x^2 = -k^2 \phi(x)$. In contrast, the application of the second-order discrete Laplacian operator for the uniform mesh of Fig.~\ref{fig:1dGrid} yields
\begin{equation}
    \frac{\phi(x-h)  -2 \phi(x) + \phi(x+h)}{h^2} = \frac{e^{-ikh} + e^{+ikh} -2}{h^2} e^{ikx} =
    \frac{2 \, \text{cos}(kh) - 2}{h^2} \phi(x).
\end{equation}
That is, the discrete Laplacian operator reduces the actual wavenumber $k$ into a modified wavenumber $\sqrt{ (2 - 2\, \text{cos}(kh))/h^2}$. A similar analysis can be performed for the D4 operator. Namely, with $\myD4(h) = 7/6 \myD2(h) - 1/6 \myD2(2h)$, the application of the discrete $\myD4$ operator on the single wave fuction $\phi(x) = e^{ikx}$ provides the following result
\begin{equation}
    \myD4 (\phi(x)) =
    \li \frac{7}{6}\frac{2 \, \text{cos}(kh) - 2}{h^2}  -
     \frac{1}{6}\frac{2 \, \text{cos}(2kh) - 2}{(2h)^2} \re  \phi(x).
\end{equation}
Therefore, the discrete $\myD4$ operator provides a modified wavenumber which is equal to $ \sqrt{ \li \frac{7}{6}(2 - 2 \, \text{cos}(kh))/{h^2}  -
     \frac{1}{6}(2 - 2 \, \text{cos}(2kh))/(2h)^2 \re }$.
Figure~\ref{fig:modifiedk} shows the modified wavenumbers for the D2 and D4 operators as  function of $kh$. From this figure, it can be concluded that the modified wavenumber of the D4 operator is closer to the analytical wavenumber than the modified wavenumber of the D2 operator. This means that the D4 operator resolves the physical dissipation better than the D2 operator, which was observed  in the examples presented in section~\ref{sec:energydissipation}.

For distorted meshes with cell stretching, non-orthogonality, and skewness, the analysis becomes more complex, because the sinus/cosinus are not eigenvectors of the discrete Laplacian operator anymore. Therefore, a similar analysis for distorted meshes is out of the scope of this paper.

\begin{figure}[ht]
	\centering
	\includegraphics[width=0.6 \textwidth]{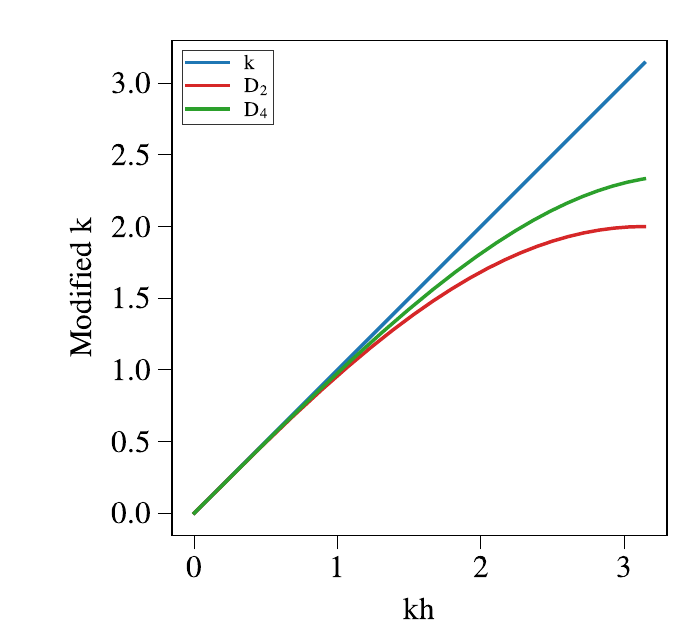}
	\caption{Modified wave number as function of $kh$.}
	\label{fig:modifiedk}
\end{figure}

} 


\newpage
\bibliographystyle{elsarticle-harv}
\bibliography{references}

\end{document}